%% file: longpaper.tex
\documentclass[aps,prd,preprintnumbers,amssymb,nobibnotes,nofootinbib,longbibliography,superscriptaddress]{revtex4-2}
\usepackage[utf8]{inputenc}
\usepackage[T1]{fontenc}

\usepackage{amsmath, amsfonts, amssymb}
\usepackage{dsfont}
\usepackage{mathtools}
\usepackage{physics}

\usepackage{graphicx}
\usepackage{svg}

\usepackage{booktabs}
\usepackage{array}

\usepackage{enumitem}

\graphicspath{{figures/}}

\usepackage[%
]{hyperref}

\hypersetup{%
    colorlinks=true,%
    allcolors={blue!60!black}%
}

\usepackage{cleveref}

\newtheorem{example}{Example}

\newcommand{\gammacircle}[1]{
  \ensuremath{%
  \gamma_{\circlearrowleft}^{#1}
}}

\newcommand{\dlog}[1]{\dd{\log{#1}}}

\newcommand{\ii}{\,\mathrm{i}\,}
\newcolumntype{C}[1]{>{\centering\arraybackslash}p{#1}}
\newcommand{\red}[1]{{\color{red}{#1}}}

\newcommand{\IefactE}{\vb{I}_{\mathcal{E}}}

\newcommand{\JomnC}{\vb{J}_{\mathcal{C}}}

\begin{document}

\title{How to choose a good rational basis for elliptic Feynman integrals?}

\author{E.~Chaubey}
\affiliation{Bethe Center for Theoretical Physics, Universität Bonn, 53115 Bonn, Germany}
\author{V.~Sotnikov}
\affiliation{Institute of Physics, Johannes Gutenberg University Mainz, Staudinger Weg 7, 55099 Mainz, Germany}

\date{\today}

\preprint{BONN-TH-2026-18}
\preprint{MITP-26-044}

\begin{abstract}
Representing multi-loop scattering amplitudes as linear combinations of multivalued transcendental functions with process-dependent rational coefficients has long been understood to be advantageous.
In general, these transcendental functions satisfy a system of differential equations with coupled homogeneous blocks.
When these coupled blocks can be removed through algebraic basis transformations, the relation between the rational and algebraic bases is universal and minimal.
Here, we ask whether an analogous universal and minimal relation exists when decoupling requires transformations involving complete elliptic integrals.
We elaborate on the method proposed in Ref.~\cite{Chaubey:2025adn}, in which we suggested that a basis constructed using an elliptic generalization of leading singularities may provide an answer to this question.
We extend this analysis to the off-diagonal blocks of the rational differential equations satisfied by the bases obtained through this construction.
We find that the $\epsilon$ dependence of these blocks can be organized into a universal structure that is preserved under the decoupling transformation used to express the solutions in terms of iterated integrals.

\end{abstract}

\maketitle 
\tableofcontents


\section{Introduction}

An analytic understanding of scattering amplitudes has long been essential both for precision high-energy phenomenology and for uncovering the formal structure of amplitudes and quantum field theories more broadly.
Beyond leading order in perturbation theory, scattering amplitudes develop a rich non-analytic structure, including branch cuts associated with the production of intermediate states, while their poles encode universal factorization properties.
A fruitful approach to their analytic computation is to express amplitudes in a basis of transcendental functions that capture this general analytic structure and originate from Feynman integrals, multiplied by rational coefficients that encode process-specific information such as the interactions and spins of the external states.

Maintaining this conceptual and practical separation has proved particularly valuable in multiscale calculations.
With a suitable choice of transcendental functions, modern analytic reconstruction and bootstrap techniques \cite{Peraro:2016wsq,vonManteuffel:2014ixa,Badger:2023mgf,Abreu:2018zmy,DeLaurentis:2022otd,DeLaurentis:2025dxw,Carrolo:2025agz,Henn:2018cdp} can exploit the relative simplicity of the rational coefficients, even when the intermediate stages of the computation are considerably more complicated. 
This strategy has enabled many state-of-the-art results \cite{Agarwal:2021vdh, Agarwal:2023suw, Badger:2024mir, Badger:2025ljy, DeLaurentis:2023izi, DeLaurentis:2023nss, DeLaurentis:2026brm}.

Crucially, making this simplicity manifest depends on a suitable choice of transcendental function basis.
One of the most powerful approaches to its construction is to study the differential equations (DEs) satisfied by Feynman integrals~\cite{Kotikov:1990kg,Kotikov:1991pm,Remiddi:1997ny,Bern:1993kr,Gehrmann:1999as}.
In particular, master-integral bases for which the dependence on the dimensional regulator $\epsilon$ factorizes from the DE are especially useful~\cite{Henn:2013pwa}.
Their solutions can be constructed order by order in $\epsilon$ as constant linear combinations of iterated integrals~\cite{Chen:1977oja}.
Under suitable conditions, these iterated integrals are linearly independent~\cite{Deneufchatel:2011yph,Duhr:2024xsy,Duhr:2025xyy} and can therefore be regarded as basis vectors of the transcendental function space in which the amplitude takes values.
The coefficients of different Feynman integrals at a given order in their $\epsilon$ expansion are generally redundant linear combinations of these basis vectors.
These redundancies can be eliminated algorithmically to construct a basis of transcendental functions~\cite{Gehrmann:2018yef,Chicherin:2020oor,Chicherin:2021dyp}.
The resulting basis functions satisfy a nilpotent DE: differentiation lowers the iteration length, with no homogeneous term at fixed length.

However, except in special cases, an $\epsilon$-factorized integral basis cannot be obtained through rational linear combinations of the Feynman integrals naturally defined by momentum-space Feynman rules.
Consequently, one cannot in general define a transcendental function basis through constant linear combinations of iterated integrals while maintaining a clean separation between the rational and multivalued parts of the amplitude.
From this perspective, function bases in which scattering amplitudes have rational coefficients must instead satisfy rational DEs with coupled homogeneous blocks.
Removing these homogeneous blocks requires introducing solutions of the corresponding homogeneous DEs into the basis transformation.
Although Feynman integrals are naturally multivalued mathematical objects, their physical values are fixed unambiguously by boundary data and the causal prescription of the quantum field theory.
By contrast, the normalization and analytic continuation of the auxiliary homogeneous solutions are not fixed by these physical requirements.
Introducing such solutions into the basis transformation therefore gives rise to unphysical ambiguities analogous to gauge dependence.
Here we discuss how these ambiguities can be quantified by the monodromy group of the homogeneous DEs (see also Ref.~\cite{Lee:2025rik} for a general discussion of the monodromy of dimensionally regulated Feynman integrals),
and emphasize that clean separation between the rational and multivalued parts of a scattering amplitude requires that the basis functions are invariant under the action of this group.

This perspective is also relevant to what has so far been the most broadly successful general approach to the numerical evaluation of complicated multiscale Feynman integrals: solving their DEs directly.
This includes the analytic continuation of local series expansions~\cite{Lee:2017qql,Moriello:2019yhu,Hidding:2020ytt,Armadillo:2022ugh}, numerical quadrature~\cite{Boughezal:2007ny,Czakon:2008zk,Mandal:2018cdj,PetitRosas:2025xhm,Badger:2025ljy}, and the construction of suitable interpolating representations~\cite{Liu:2026hdp,Liu:2026cpf,Abreu:2026vxw,Chen:2026pyw}.
Notably, neither $\epsilon$ factorization nor the nilpotence of the DEs for the transcendental functions is by itself a prerequisite for many of these approaches, which can be applied directly to rational DEs.
Retaining a rational DE avoids introducing auxiliary multivalued functions and their associated monodromy ambiguities into the numerical problem.
If a non-rational transformation is used instead, its auxiliary branches and monodromy dependence require a consistent choice of ``gauge fixing'' (see, for example, Refs.~\cite{Baur:2026zlw,Czakon:2026tog}).

Function bases satisfying rational DEs therefore provide an important interface between iterated-integral representations, whose kernels encode the underlying geometry, and function bases suitable for representing scattering amplitudes with rational coefficients.
This raises the central question of how such rational bases should be chosen.

The situation is particularly transparent and well understood when the homogeneous DE can be decoupled by algebraic transformations.
In this case, the relation between a rational basis and an $\epsilon$-factorized algebraic basis is simple and universal: the required transformation is determined by the algebraic solutions of the homogeneous system at leading order in $\epsilon$.
Although the corresponding algebraic factors carry non-trivial monodromy, transforming back to the rational basis typically amounts to a simple normalization of the transcendental functions by these factors, producing monodromy-invariant combinations.

For elliptic Feynman integrals, however, the situation is considerably more involved.
The leading-order homogeneous system may have genuinely transcendental solutions involving complete elliptic integrals.
Most existing approaches focus on constructing an $\epsilon$-factorized basis through transformations involving such transcendental functions~\cite{Adams:2018bsn,Dlapa:2022wdu,Gorges:2023zgv,Chen:2025hzq,Yang:2025ofz,Duhr:2025xyy,e-collaboration:2025frv,Bree:2025tug,Bree:2026qww,Forner:2026vby,Dlapa:2022wdu}.
These transformations can be considerably more complicated than in the algebraic case and generally depend simultaneously on the kinematics, $\epsilon$, and a set of complete elliptic integrals.
Moreover, additional transcendental functions may be required once the off-diagonal blocks coupling different sectors are taken into account \cite{e-collaboration:2025frv,Duhr:2025lbz,Becchetti:2025oyb,Duhr:2025xyy,Gorges:2023zgv,Adams:2018kez}.
From the perspective of a clean separation between the rational and multivalued parts of an amplitude, this raises an important question: \emph{is there a preferred rational basis for elliptic Feynman integrals that provides a natural starting point for constructing transcendental function bases for scattering amplitudes?}

Several recent developments point towards useful structures in this direction. 
The geometric organization of Refs.~\cite{e-collaboration:2025frv,Bree:2025tug} leads, on the maximal cut, to DEs with a controlled Laurent-polynomial dependence on $\epsilon$.
Complementary criteria for selecting rational bases include Lee's symmetric form and finite or quasi-finite integral bases~\cite{Lee:2018jsw,Chetyrkin:2006dh,vonManteuffel:2014qoa,Lee:2019wwn,DeAngelis:2025agn}.
Practical choices of rational bases with simplified $\epsilon$ dependence, made without explicitly constructing them from the underlying elliptic geometry, have also proved useful in multiscale applications~\cite{Badger:2024fgb,Becchetti:2025qlu}.

In Ref.~\cite{Chaubey:2025adn}, we proposed an approach for choosing a rational basis inspired by an elliptic generalization of leading singularities suggested in refs.~\cite{Bourjaily:2020hjv,Bourjaily:2021vyj} (see also \cite{Chen:2025hzq} for a related approach).
The underlying idea parallels the familiar observation that integrals with $\dd\log$ integrands and unit leading singularities provide good candidates in the algebraic case~\cite{Cachazo:2008vp,Arkani-Hamed:2010pyv}.
In the elliptic case, candidate integrals are instead constructed by matching their integrands to algebraic one-forms naturally selected by the cohomology of the elliptic curve.
The resulting homogeneous DEs were found to be particularly simple, with a dependence on $\epsilon$ that is at most linear.
In addition, the coupled DE block at leading order in $\epsilon$ minimally encodes the transcendental transformation required to decouple the homogeneous system and express its solutions in terms of iterated integrals.

The analysis of Ref.~\cite{Chaubey:2025adn}, however, was largely restricted to the homogeneous DEs, which are determined by the maximal cut.
For applications to complete scattering amplitudes, this is not sufficient: one must also understand the off-diagonal blocks that couple a given sector to its subsectors.
This distinction is particularly important from the perspective adopted here.
A common strategy is to first $\epsilon$-factorize the homogeneous elliptic block, thereby introducing transcendental functions, and subsequently use additional transformations to $\epsilon$-factorize the off-diagonal blocks.
If the goal is instead to construct a rational basis that leads to monodromy-invariant transcendental functions, the order of these operations is reversed: the subsector dependence should first be simplified within the rational basis, and only afterwards should the minimal transcendental transformation required to decouple the elliptic homogeneous block be introduced.

In this work, we develop this perspective further.
We first revisit in detail the relation between rational, algebraic, and elliptic bases from the viewpoint of separating rational coefficients from multivalued transcendental functions.
This makes explicit why the rationality of the integral basis is closely related to the monodromy invariance of the corresponding function basis.
We then review and elaborate on the elliptic-leading-singularity construction of Ref.~\cite{Chaubey:2025adn}, clarifying its implementation and limitations.
We also explain the origin of the integer $n$ that appears prominently in the DEs satisfied by these bases.
Our main new result concerns the structure of the off-diagonal DE blocks.
We find that, for rational bases constructed according to our prescription, the $\epsilon$ dependence of these blocks can be organized into a simple form that is preserved under the transformation used to decouple the homogeneous elliptic block.
This provides evidence that the simplicity previously observed on the maximal cut extends to the coupling with subsectors.

The resulting picture suggests a systematic strategy for multiscale elliptic Feynman integrals.
A geometry-inspired rational basis with controlled $\epsilon$ dependence identifies the subspace of monodromy-invariant transcendental functions appropriate for a clean separation between the rational and multivalued parts of the amplitude.
An $\epsilon$-factorized basis, in turn, allows the solutions to be expressed in terms of linearly independent iterated integrals and thereby makes it possible to construct an explicit basis for this subspace.

The remainder of the article is organized as follows.
In \cref{sec:function-bases}, we introduce our notation and discuss the relation between rationality, $\epsilon$ factorization, and monodromy invariance, beginning with rational and algebraic systems before turning to elliptic Feynman integrals.
In \cref{sec:recap-method}, we review and extend the construction of rational bases from elliptic leading singularities and explain the origin of the integer $n$.
In \cref{sec:inh-de}, we study the off-diagonal DE blocks and derive their restricted dependence on $\epsilon$.

\section{Transcendental function bases for scattering amplitudes}
\label{sec:function-bases}

\newcommand{\epmin}{0}
\newcommand{\epmax}{\epsilon_{\max}}

In this section, we discuss the relation between rational Feynman integral bases, $\epsilon$-factorized DEs, and transcendental function bases for scattering amplitudes.
We focus in particular on the non-rational transformations required to remove homogeneous DE blocks and on the associated monodromy, beginning with algebraic systems before turning to elliptic Feynman integrals.
Although the individual subjects of this discussion are well known in principle, their implications for separating the rational and multivalued parts of an amplitude are not usually made explicit.
We therefore review them to establish our notation, emphasize the points needed in the remainder of this paper, and provide a more detailed technical motivation for the construction introduced in Ref.~\cite{Chaubey:2025adn}.

To calculate scattering amplitudes one first reduces them to a basis of Feynman integrals, referred to as master integrals $\vb{I}(\vb*{s},\epsilon)$, such that
\begin{equation}
\label{eq:amp-mi-generic}
\mathcal{A}(\vb*{s},\epsilon)
=
\sum_i
c_i(\vb*{s},\epsilon)\,
\vb{I}_i(\vb*{s},\epsilon),
\end{equation}
with coefficients $c_i(\vb*{s},\epsilon)$.
The master integrals $\vb{I}_i(\vb*{s},\epsilon)$ are defined as linear combinations of Feynman integrals organized into \emph{families}
\begin{equation}
  \label{eq:loop-momentum-rep}
  G^{\vb{i}}_{\text{f}}
  = 
  \int \prod_{r} \dd^{d}{\ell_r}\,
  \frac{1}{\vb*{\rho}^{\vb{i}}_\text{f} },
  \qquad
  \vb*{\rho}^{\vb{i}}_{\text{f},j}
  =
  \prod_j \rho_{\text{f},j}^{\,i_j}  \,,
\end{equation}
where $\rho_{\text{f},j}$ are the inverse propagators, the index $\text{f}$ denotes different sets of propagators, and $i_j$ are integer indices, which may be positive, zero, or negative.
The integrals are regulated using dimensional regularization by analytically continuing the space-time dimension to $d = d_0 - 2 \epsilon$ dimensions with integer $d_0$.

We collectively denote the Mandelstam invariants and mass scales by $\vb*{s}$.
These variables are natural for describing the singularity structure of Feynman integrals because the Landau singular locus \cite{Correia:2025wtb, Dlapa:2023cvx, Fevola:2023fzn, Fevola:2023kaw, Helmer:2024wax, Helmer:2025ljj, Mizera:2021icv} is defined by polynomial equations in the Mandelstam invariants and squared masses.
It is therefore useful to consider master-integral bases over $\mathbb{Q}(\vb*{s},\epsilon)$ and to keep track of whether the coefficients $c_i(\vb*{s},\epsilon)$ multiplying the master integrals in \cref{eq:amp-mi-generic} are rational in $\vb*{s}$ and $\epsilon$.\footnote{
  More precisely, the coefficients are rational functions in spinor brackets. This distinction is not important for the present discussion.
}
This provides a useful separation between the rational singularities carried by the coefficients and the nontrivial branch structure contained in the master integrals.

The master integrals satisfy a system of differential equations (DEs),
\begin{equation}
\label{eq:de-generic}
\dd \vb{I}(\vb*{s},\epsilon)
=
M(\vb*{s},\epsilon)\,
\vb{I}(\vb*{s},\epsilon),
\end{equation}
where $M(\vb*{s},\epsilon)$ is a matrix-valued differential one-form, also known as connection. 
We expand $M$ around  $\epsilon=0$ as
\begin{equation}
\label{eq:de-epsilon-expansion}
M(\vb*{s},\epsilon)
=
\sum_{k=k_\text{min}}^{\infty}
\epsilon^k M^{(k)}(\vb*{s}) \,,
\end{equation}
and unless otherwise stated we assume that $k_\text{min}=0$.

For an integral with index vector $\vb{i}$, its sector is the set $S=\{j\mid i_j>0\}$ of propagators appearing in the denominator. A proper subsector is obtained by setting one or more positive indices to zero or negative values, while the top sector contains all propagators of the family that are allowed to have positive indices. Grouping the master integrals by sector and ordering the sectors according to the sector hierarchy makes the DE matrix block triangular.
Its diagonal blocks govern the homogeneous systems for master integrals within each sector, while its off-diagonal blocks encode inhomogeneous contributions from subsectors.
We will refer to the diagonal blocks of $M^{(0)}$ as the leading-order (LO) homogeneous blocks.

Cutting a propagator means taking the residue at the corresponding inverse-propagator pole $\rho_j=0$. The maximal cut of a sector is obtained by simultaneously cutting all propagators that define it. Since every proper subsector lacks at least one of these poles, its contribution vanishes on the maximal cut. The maximal cut therefore isolates the corresponding diagonal block of the DE matrix. Releasing one or more cuts allows the relevant subsector contributions to survive and provides access to the off-diagonal blocks.

Given the values of the master integrals at a point $\vb*{s}_1$, the DEs can be used to transport them to another point $\vb*{s}_2$ along a path $\gamma$ that avoids the singular locus:
\begin{equation}
\label{eq:path-exp-sol}
  \vb{I}(\vb*{s}_2,\epsilon)
  =
  U_\gamma(\vb*{s}_1,\vb*{s}_2)\,
  \vb{I}(\vb*{s}_1,\epsilon),
  \qquad
  U_\gamma(\vb*{s}_1,\vb*{s}_2)
  =
  \mathcal{P}\exp\qty(
    \int_\gamma M(\vb*{s},\epsilon)
  )  \,,
\end{equation}
where $\mathcal{P}\exp$ is path-ordered exponential.
Let $\Phi(\vb*{s},\epsilon)$ be a fundamental matrix of \cref{eq:de-generic} whose columns correspond to a basis of the solutions of \cref{eq:de-generic}.
Analytically continuing $\Phi_\gamma(\vb*{s}_1,\epsilon)$ from $\vb*{s}_1$ to $\vb*{s}_2$ along $\gamma$ gives $\Phi_\gamma(\vb*{s}_2,\epsilon)$, and the corresponding transport matrix is
\begin{equation}
  U_\gamma(\vb*{s}_1,\vb*{s}_2)
  =
  \Phi_\gamma(\vb*{s}_2)\,
  \Phi(\vb*{s}_1)^{-1},
\end{equation}
where the dependence of $U_\gamma$ and $\Phi$ on $\epsilon$ is implicit in the last equation.

\begin{figure}[ht]
  \centering
  \includegraphics[width=0.25\textwidth]{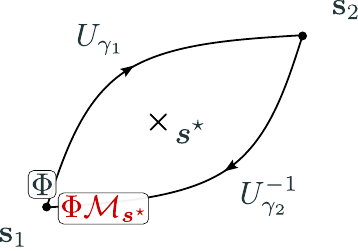}
  \caption{
    A schematic illustration of how transporting the fundamental matrix of a differential equation along a loop that encloses a singular point
    results in a transformation of the fundamental matrix by the monodromy matrix.
  }
  \label{fig:monodromy}
\end{figure}

The integrability of \cref{eq:de-generic} implies that the transport matrix is invariant under continuous deformations of $\gamma$ with fixed endpoints, provided that the deformation does not cross the singular locus of $M(\vb*{s},\epsilon)$. Consequently, $U_\gamma(\vb*{s}_1,\vb*{s}_2)$ depends only on the homotopy class of $\gamma$. This dependence is quantified by the \emph{monodromy}.
Let $\gamma_1$ and $\gamma_2$ be two paths from $\vb*{s}_1$ to $\vb*{s}_2$ such that the relative path
$\gammacircle{\vb*{s}_\star}=\gamma_2^{-1}\circ\gamma_1$
is a closed loop based at $\vb*{s}_1$ that encircles a singular surface at $\vb*{s}_\star$. The corresponding transport matrices satisfy
\begin{equation}
\label{eq:monodromy-def}
  U_{\gamma_1}
  =
  U_{\gamma_2}\,
  \mathcal{M}_{\vb*{s}_\star},
  \qquad
  U_{\gammacircle{\vb*{s}_\star}}\,\Phi
  =
  \Phi\,\mathcal{M}_{\vb*{s}_\star},
\end{equation}
where all arguments are implicit and $\mathcal{M}_{\vb*{s}_\star}$ is the \emph{monodromy matrix}.
This is illustrated in \cref{fig:monodromy}.
The physical branch of the Feynman integrals is fixed by their value at a base point, together with analytic continuation along a homotopy class compatible with Feynman's $\ii\delta$ prescription. This prescription determines how singularities encountered on the physical sheet are bypassed. More generally, we call a singular surface of $M(\vb*{s},\epsilon)$ spurious for the physical solution if the corresponding monodromy acts trivially on the vector of Feynman integrals. The monodromy of the full fundamental matrix may nevertheless be nontrivial around such a surface.

Expanding the transport equation \cref{eq:path-exp-sol} in $\epsilon$ yields the master integrals to any required order $\epmax$,
\begin{equation}
\label{eq:int-eps-expansion}
  \vb{I}(\vb*{s},\epsilon)
  =
  \sum_{k=\epmin}^{\epmax}
  \epsilon^k\,\vb{I}^{(k)}(\vb*{s})
  +
  \order{\epsilon^{\epmax+1}}  \,,
\end{equation}
where we normalize all integrals such that they start from order $\epsilon^0$ (which may be vanishing).
The functions appearing in the expansion coefficients $\vb{I}^{(k)}$ need not be linearly independent over $\mathbb{Q}(\vb*{s})$. At a fixed truncation order, there may therefore be nontrivial relations
\begin{equation}
\label{eq:linear-dependencies}
  \sum_{k=\epmin}^{\epmax}
  \sum_j
  r_j^{(k)}(\vb*{s})\,
  \vb{I}^{(k)}_j(\vb*{s})
  =
  0,
  \qquad
  r_j^{(k)}(\vb*{s})\in\mathbb{Q}(\vb*{s}).
\end{equation}
We eliminate these redundancies by choosing a basis of transcendental functions,
\begin{equation}
\label{eq:trans-basis}
  f_i(\vb*{s})
  =
  \sum_{k=\epmin}^{\epmax}
  \sum_j
  r_{ij}^{(k)}(\vb*{s})\,
  \vb{I}^{(k)}_j(\vb*{s}),
\end{equation}
that is linearly independent over $\mathbb{Q}(\vb*{s})$ and spans all functions appearing in \cref{eq:int-eps-expansion}.

Reexpressing the integral coefficients in this basis and inserting them into \cref{eq:amp-mi-generic} gives
\begin{equation}
\label{eq:ampl-basis}
  \mathcal{A}^{(L)}(\vb*{s},\epsilon)
  =
  \sum_{k=-2L}^{\epsilon_\text{amp} }
  \epsilon^k
  \sum_i
  c_i^{(k)}(\vb*{s})\,
  f_i(\vb*{s})
  +
  \order{\epsilon^{\epsilon_\text{amp} +1}}.
\end{equation}
Here, the lower limit $-2L$ is the maximal infrared pole expected for an $L$-loop amplitude. The truncation order $\epsilon_\text{amp}$ is usually chosen as the minimum required to obtain the finite observable of interest after combining all relevant perturbative contributions.

A key problem in analytic amplitude computations is to choose the transcendental functions $f_i(\vb*{s})$ in \cref{eq:ampl-basis} such that their coefficients $c_i(\vb*{s})$ are as simple as possible. The two basic requirements are
\begin{enumerate}
  \item[\rm(i)] that the multivaluedness of the Feynman integrals is fully contained in $f_i(\vb*{s})$, while their coefficients are rational;
  \item[\rm(ii)] that the functions $f_i(\vb*{s})$ are linearly independent.
\end{enumerate}
These properties are essential for applying modern analytic reconstruction techniques~\cite{Peraro:2016wsq,vonManteuffel:2014ixa,Badger:2023mgf,Abreu:2018zmy,DeLaurentis:2022otd,DeLaurentis:2025dxw} to the rational coefficients and for bootstrap approaches~\cite{Carrolo:2025agz}. 
A suitable choice of basis is also a prerequisite for existence of compact representations of rational coefficients which reduces numerical cancellations and facilitates the efficient and stable numerical evaluation of amplitudes for phenomenological applications~\cite{Gehrmann:2018yef,Chicherin:2020oor,Chicherin:2021dyp,Abreu:2023rco}.

\subsection{Rational case}
\label{sec:rational-case}

The simplest case in which the rational and transcendental parts can be separated explicitly is provided by a \emph{rational canonical basis}. Let $\vb{L}(\vb*{s},\epsilon)$ denote a reference basis produced by the IBP reduction. The rational case is defined by the existence of an invertible transformation
\begin{equation}
\label{eq:rat-transformation}
  \vb{I}(\vb*{s},\epsilon)
  =
  T(\vb*{s},\epsilon)\,
  \vb{L}(\vb*{s},\epsilon),
  \qquad
  T_{ij}(\vb*{s},\epsilon)
  \in
  \mathbb{Q}(\vb*{s},\epsilon),
\end{equation}
such that the DE for $\vb{I}(\vb*{s},\epsilon)$ takes the canonical form~\cite{Henn:2013pwa}
\begin{equation}
\label{eq:de-canonical}
  \dd\vb{I}(\vb*{s},\epsilon)
  =
  \epsilon\,M(\vb*{s})\,
  \vb{I}(\vb*{s},\epsilon),
  \qquad
  M(\vb*{s})
  =
  \sum_i M_i\,\omega_i(\vb*{s})
  =
  \sum_i M_i\,\dd\log W_i(\vb*{s}),
\end{equation}
where the $M_i$ are constant matrices with rational entries and the letters
$W_i(\vb*{s})$ are rational functions of the kinematic variables.  Thus, the
dependence on $\epsilon$ is factorized and the connection is expressed in
terms of logarithmic one-forms with rational arguments. Systematic algorithms
exist for finding such a rational canonical basis whenever it
exists~\cite{Lee:2014ioa,Meyer:2017joq}.

The usefulness of the canonical form follows directly from its path-ordered
solution. Substituting \cref{eq:de-canonical} into \cref{eq:path-exp-sol} gives
\begin{equation}
\label{eq:pexp-ep-fact}
  U_\gamma(\vb*{s}_1,\vb*{s}_2)
  =
  \mathds{1}
  +
  \sum_{k=1}^{\infty}
  \epsilon^k
  \sum_{i_1,\ldots,i_k}
  M_{i_k}\cdots M_{i_1}\,
  I_\gamma(\omega_{i_1},\ldots,\omega_{i_k}),
\end{equation}
where
\begin{equation}
\label{eq:iint}
  I_\gamma(\omega_{i_1},\ldots,\omega_{i_k})
  =
  \int_{0<t_1<\cdots<t_k<1}
  \prod_{r=1}^{k}
  w_{i_r}(t_r)\,\dd t_r,
  \qquad
  \gamma^*\omega_i=w_i(t)\,\dd t,
\end{equation}
are Chen iterated integrals~\cite{Chen:1977oja}. Here
$\gamma\colon[0,1]\to\mathcal{K}$ is a path in kinematic space $\vb*{s} \in \mathcal{K}$ with
$\gamma(0)=\vb*{s}_1$ and $\gamma(1)=\vb*{s}_2$,
and $\gamma^*\omega_i$ is a pull-back onto the path $\gamma$.  Consequently, the coefficient
of $\epsilon^k$ in the transport matrix is expressed in terms of iterated
integrals of length $k$. After including the $\epsilon$ expansion of the
boundary values, the coefficient at order $\epsilon^k$ involves iterated
integrals of length at most $k$.

Under suitable conditions on the logarithmic one-forms, an independent set
of iterated integrals associated with distinct sequences of one-forms (\emph{words}) is linearly independent
over the field of rational functions~\cite{Deneufchatel:2011yph}. 
Because the matrices $M_i$ have rational entries, the $\epsilon$-series coefficients of
the canonical master integrals are generated by $\mathbb{Q}$-linear
combinations of iterated integrals, once the boundary constants are included
among the transcendental generators. The reduction to the basis in
\cref{eq:trans-basis} can therefore be performed by linear algebra over
$\mathbb{Q}$ rather than over $\mathbb{Q}(\vb*{s})$.

The full space of integrable words over the alphabet $\{\omega_i\}$ grows
rapidly with the length and is generally much larger than the subspace
actually required by the master integrals. A practical approach is therefore
to construct only the combinations appearing in the $\epsilon$ expansion of
the master integrals and then determine a basis of their span by
$\mathbb{Q}$-linear algebra~\cite{Gehrmann:2018yef,Chicherin:2020oor,Chicherin:2021dyp}.
This role of the rational canonical basis is important for the discussion
below: it makes the separation in \cref{eq:ampl-basis} explicit, with the
multivalued dependence contained in $\mathbb{Q}$-linear combinations of
iterated integrals and the amplitude coefficients given by rational
functions.

Finally, let us recall how \emph{leading singularities} (LS) \cite{Cachazo:2008vp}  provide a useful guiding principle for constructing rational canonical bases. Suppressing the integral-family label $\mathrm{f}$, an algebraic parametrization $\mathcal{P}$ maps each integral to a differential $k$-form in the integration variables $\vb*{z}=(z_1,\ldots,z_k)$,
\begin{equation}
\label{eq:parametrization}
  \mathcal{P}\qty(G^{\vb{i}})
  =
  g^{\vb{i}}(\vb*{s},\vb*{z})\,
  \dd^k{\vb*{z}}
  +
  \order{\epsilon}  \,,
\end{equation}
where $\dd^k{\vb*{z}}\equiv\dd z_1\wedge\cdots\wedge\dd z_k$. We call the differential form
$g^{\vb{i}}(\vb*{s},\vb*{z})\,\dd^k{\vb*{z}}$
the integrand of $G^{\vb{i}}$ in fixed $d_0$ dimensions. Any suitable parametrization may be used. In the following, both the Baikov parametrization~\cite{Baikov:1996iu,Frellesvig:2017aai,Bosma:2017ens} and direct momentum-space parametrizations~\cite{Henn:2020lye} are employed.

Given a finite set of integrals, consider the ansatz
\begin{equation}
\label{eq:ansatz}
  \mathcal{I}[n]
  =
  \sum_{\vb{i}}
  n_{\vb{i}}(\vb*{s})\,
  G^{\vb{i}}
  \;\xrightarrow{\mathcal{P}}\;
  \underbrace{
  \sum_{\vb{i}}
  n_{\vb{i}}(\vb*{s})\,
  g^{\vb{i}}(\vb*{s},\vb*{z})\,
  \dd^k{\vb*{z}}
  }_{\displaystyle \Omega[n](\vb*{s},\vb*{z})}
  ,
  \qquad
  n_{\vb{i}}(\vb*{s})\in\mathbb{Q}(\vb*{s}),
\end{equation}
where $n$ denotes the collection of unknown coefficients $n_{\vb{i}}$.

The construction then proceeds in three steps. First, the coefficients are constrained such that the integrand $\Omega[n]$ admits a representation as a logarithmic $k$-form~\cite{Cachazo:2008vp,Bern:2014kca,Herrmann:2019upk,Henn:2020lye,Bourjaily:2020hjv,Dlapa:2021qsl},
\begin{equation}
\label{eq:dlog-integrands}
  \Omega[n](\vb*{s}, \vb*{z})
  =
  \sum_{j=1}^{n_{\text{LS}}}
  l_j[n](\vb*{s})\,
  \omega_j(\vb*{s},\vb*{z}),
  \qquad
  \omega_j
  =
  \dlog{\alpha_{j,1}}
  \wedge\cdots\wedge
  \dlog{\alpha_{j,k}}.
\end{equation}
The LS are then multidimensional residues
\begin{equation}
\label{eq:localized-integrand}
  l_j[n](\vb*{s})
  =
  \frac{1}{(2\pi\ii)^k}
  \int_{\otimes_{r=1}^k\gammacircle{\alpha_{j,r}}}
  \Omega[n](\vb*{s},\vb*{z}).
\end{equation}

Second, the solution space of the logarithmic constraints is diagonalized with respect to an independent set of LS. A set of solutions $n^{(a)}$ is chosen such that
\begin{equation}
\label{eq:LS-diagonalization}
  \widehat{I}_a
  =
  \mathcal{I}[n^{(a)}],
  \qquad
  l_j[n^{(a)}](\vb*{s})
  =
  \lambda_a(\vb*{s})\,
  \delta_j^{\,a}.
\end{equation}
Thus, each candidate $\widehat{I}_a$ has only one non-vanishing LS $\lambda_a(\vb*{s})$.

Third, each candidate is divided by its remaining LS:
\begin{equation}
\label{eq:LS-normalization}
  I_a
  =
  \frac{\widehat{I}_a}{\lambda_a(\vb*{s})},
  \qquad
  \frac{1}{(2\pi\ii)^k}
  \int_{\otimes_{r=1}^k\gammacircle{\alpha_{j,r}}}
  \Omega[n_a]
  =
  \delta_j^{\,a}.
\end{equation}
In the rational case, $\lambda_a(\vb*{s})$ is rational, so the normalization preserves the rationality of the basis transformation. The resulting integrals have unit LS and provide natural candidates for a rational canonical basis.

In multiscale computations however, it is rather an exception that such bases exist.
Motivated by the convenience of basis construction, let us now discuss what happens if we insist on constructing bases that satisfy $\epsilon$-factorized DE as a guidance when this is not possible to achieve by a rational transformation as in \cref{eq:rat-transformation}.

\subsection{Algebraic case}
\label{sec:algebraic-case}

\subsubsection*{Square-root normalization}

A common extension of the rational case occurs when the canonical form in
\cref{eq:de-canonical} can be reached only after allowing an \emph{algebraic}
change of basis. The simplest such case involves
square-root normalizations,
\begin{equation}
\label{eq:algebraic-basis-sqrt-simple}
  \vb{I}_{\mathrm A}(\vb*{s},\epsilon)
  =
  T_{\mathrm A}(\vb*{s})\,
  \vb{I}_{\mathrm R}(\vb*{s},\epsilon)
  =
  T_{\mathrm A}(\vb*{s})\,
  T(\vb*{s},\epsilon)\,
  \vb{L}(\vb*{s},\epsilon),
  \qquad
  T_{ij}(\vb*{s},\epsilon)
  \in
  \mathbb{Q}(\vb*{s},\epsilon),
  \qquad
  T_{\mathrm A}(\vb*{s})
  =
  \mathrm{diag}\!\left\{r_i^{-1/2}\right\},
\end{equation}
where $r_i\in\mathbb{Q}[\vb*{s}]$, with $r_i=1$ explicitly allowed.
Here $\vb{I}_{\mathrm R}$ is rationally related to the reference IBP basis,
whereas $\vb{I}_{\mathrm A}$ is the algebraically normalized canonical basis.

The origin of the square-root factors follows directly from the three-step
leading-singularity construction described in \cref{sec:rational-case}. As in the
rational case, a rational ansatz is first constrained to have a $\dd\log$
integrand and is then diagonalized so that each candidate integral has a
single non-vanishing leading singularity. The difference arises only in the
final normalization step. After extracting a rational factor, the diagonal
leading singularity may take the form
\begin{equation}
  l_j(\vb*{s})
  =
  q_j(\vb*{s})\sqrt{r_j(\vb*{s})},
  \qquad
  q_j(\vb*{s})
  \in
  \mathbb{Q}(\vb*{s}).
\end{equation}
The rational factor can be removed while remaining within a rational
master-integral basis, but the square root cannot. Thus,
\begin{equation}
\label{eq:algebraic-ls-normalization}
  \vb{I}_{\mathrm R,j}
  =
  \frac{\widehat{\vb{I}}_{\mathrm R,j}}
       {q_j(\vb*{s})},
  \qquad
  \mathrm{LS}\!\left(\vb{I}_{\mathrm R,j}\right)
  =
  \sqrt{r_j(\vb*{s})},
  \qquad
  \vb{I}_{\mathrm A,j}
  =
  \frac{\vb{I}_{\mathrm R,j}}
       {\sqrt{r_j(\vb*{s})}},
  \qquad
  \mathrm{LS}\!\left(\vb{I}_{\mathrm A,j}\right)
  =
  1.
\end{equation}
The rational diagonalization and normalization construct
$T(\vb*{s},\epsilon)$, while the residual square-root normalization is
implemented by $T_{\mathrm A}(\vb*{s})$. Integrals with rational leading
singularities are included by setting $r_j=1$.

Since $\vb{I}_{\mathrm A}$ satisfies the canonical DE, the rational basis
obeys
\begin{equation}
\label{eq:de-rational-sqrt}
  \dd\vb{I}_{\mathrm R}(\vb*{s},\epsilon)
  =
  \left[
    -\dd\log T_{\mathrm A}(\vb*{s})
    +
    \epsilon\,M_{\mathrm R}(\vb*{s})
  \right]
  \vb{I}_{\mathrm R}(\vb*{s},\epsilon),
  \qquad
  M_{\mathrm R}
  =
  T_{\mathrm A}^{-1}\,M\,T_{\mathrm A}.
\end{equation}
Because $\vb{I}_{\mathrm R}$ is rationally related to the IBP basis, its DE
is rational. In particular, $M_{\mathrm R}$ is a rational matrix-valued
one-form, although its representation inherited from the canonical system
may involve square-root factors multiplying $\dd\log$-forms with algebraic
arguments. These factors combine to give rational entries in
\cref{eq:de-rational-sqrt}. The LO homogeneous block is diagonal, with
entries $\frac{1}{2}\dd\log r_i$.

The physical solution of \cref{eq:de-rational-sqrt} along a path is fixed by
its value at a base point and by choosing the path's homotopy class compatible with
Feynman's $\ii\delta$ prescription. It can be transported using numerical
quadrature~\cite{Boughezal:2007ny,Czakon:2008zk,Mandal:2018cdj,PetitRosas:2025xhm,Badger:2025ljy}
or local series
expansions~\cite{Lee:2017qql,Moriello:2019yhu,Hidding:2020ytt,Armadillo:2022ugh}.
For direct numerical transport, the non-vanishing LO homogeneous block poses
no conceptual difficulty.

It does, however, prevent the path-ordered exponential from truncating to
finite iterated-integral length at fixed order in $\epsilon$, since
arbitrarily many insertions of the LO connection may occur. Recovering an
$\epsilon$-factorized system therefore requires removing this homogeneous part.
For a particular square root $r_i$, the fundamental matrix $\Phi(\vb*{s})$ satisfies the DE
\begin{equation}
\label{eq:de-rational-sqrt-hom}
  \dd\Phi(\vb*{s})
  =
  \frac{\dd r_i}{2 r_i}\,
  \Phi(\vb*{s}),
\end{equation}
The solution is $\Phi=C\sqrt{r_i}$ with constant $C$, and only \cref{eq:de-rational-sqrt-hom} is required to remove the homogeneous block.
Unlike the physical solution of \cref{eq:de-rational-sqrt}, the normalization
and branch of $\Phi$ are not fixed by physical boundary conditions and may
be chosen arbitrarily, provided the choice is made consistently.
This branch ambiguity is characterized by the monodromy of the homogeneous DE. 
Analytic continuation around a small loop
$\gammacircle{r_i=0}$ gives
\begin{equation}
\label{eq:monodromy-sqrt}
  U_{\gammacircle{r_i=0}}\Phi
  =
  \exp\!\left(
    \frac{1}{2}
    \oint_{\gammacircle{r_i=0}}
    \frac{\dd r_i}{r_i}
  \right)\Phi
  =
  -\Phi.
\end{equation}
The non-trivial sign monodromy shows that $\Phi$ is not rational, since rational functions have trivial monodromy. 
Let us also note that the corresponding surface $r_i=0$ may correspond to a spurious singularity.

The transformation
\begin{equation}
  \vb{I}_{\mathrm A,i}
  =
  \Phi^{-1}\,
  \vb{I}_{\mathrm R,i}
\end{equation}
eliminates the LO homogeneous block and brings the DE to
$\epsilon$-factorized form. The algebraically normalized integral
$\vb{I}_{\mathrm A,i}$ consequently transforms in a non-trivial sign
representation of the monodromy group of \cref{eq:de-rational-sqrt-hom}.
This auxiliary ambiguity cancels in the physical combination
$\vb{I}_{\mathrm R,i}=\Phi\,\vb{I}_{\mathrm A,i}$.
The resulting canonical system can be solved in terms of iterated integrals
as in the rational case, now with algebraic one-forms. Within each monodromy
sector, the functions $\widetilde f_i(\vb*{s})$ can again be constructed as
$\mathbb{Q}$-linear combinations of the iterated integrals in
\cref{eq:pexp-ep-fact,eq:iint}. The monodromy sectors can equivalently be
encoded by a grading of the iterated-integral
algebra~\cite{Chicherin:2021dyp}.

Unlike in the rational case, however, the monodromy-invariant transcendental
basis that separates the rational and multivalued parts of the amplitude is
not formed by the functions $\widetilde f_i$ alone. They must be dressed by
algebraic factors as
\begin{equation}
\label{eq:algebraic-function-basis}
  f_i(\vb*{s})
  =
  \widetilde f_i(\vb*{s})\,
  r_i(\vb*{s})^{\sigma_i/2},
  \qquad
  \sigma_i\in\{0,\pm1\}.
\end{equation}
For a non-trivial square-root sector, this reduces to
$f_i=\widetilde f_i\,r_i^{\pm1/2}$,
where the sign in the exponent can be chosen to reduce the appearance of spurious poles in the rational coefficients \cite{DeLaurentis:2026brm}.

This organization of transcendental function bases has been essential for
many cutting-edge applications to multiscale scattering
amplitudes~\cite{DeLaurentis:2025dxw,Badger:2025ljy,Becchetti:2026yxl,Gehrmann:2024tds,Badger:2023xtl,DeLaurentis:2026brm,Agarwal:2023suw,DeLaurentis:2023izi,DeLaurentis:2023nss}.

\subsubsection*{Nested square roots}

Beyond the single-square-root normalization discussed above, nested square
roots may also be required to obtain an $\epsilon$-factorized
DE~\cite{FebresCordero:2023pww,Badger:2024fgb,Becchetti:2025oyb,Becchetti:2025qlu,Pozzoli:2026eiu,Aliaj:2026iny}.
Examples and detailed discussions of their appearance can be found in these
references. Here, two aspects relevant to the present discussion are
emphasized: the appearance of nested roots in the $\dd\log$ construction and
the monodromy representation carried by the resulting canonical integrals.

The essential difference from the square-root case already appears in the
diagonalization step of the leading-singularity construction. After removing
all candidates that can be diagonalized rationally, the remaining rational ansatz takes the schematic form
\begin{align}
  n_1\vb{I}_{\mathrm R,1}
  +
  n_2\vb{I}_{\mathrm R,2}
  &\xrightarrow{\mathcal P}
  l_1(\vb*{s},n_1,n_2)\,\Omega_1
  +
  l_2(\vb*{s},n_1,n_2)\,\Omega_2
  \,,
  \nonumber\\
  l_1(\vb*{s},n_1,n_2)
  &=
  n_1\sqrt{q+\sqrt r}
  +
  n_2\frac{\sqrt{q+\sqrt r}}{\sqrt r},
  \nonumber\\
  l_2(\vb*{s},n_1,n_2)
  &=
  n_1\sqrt{q-\sqrt r}
  -
  n_2\frac{\sqrt{q-\sqrt r}}{\sqrt r},
\end{align}
where $\Omega_1$ and $\Omega_2$ are logarithmic forms, while $q$ and $r$ are
polynomials in $\vb*{s}$ that need not be square-free. Diagonalization
requires
\begin{equation}
  l_2=0
  \;\Longrightarrow\;
  n_2=\sqrt r\,n_1
  \quad\text{or}\quad
  l_1=0
  \;\Longrightarrow\;
  n_2=-\sqrt r\,n_1.
\end{equation}
Thus, unlike in the previous section, the leading-singularities cannot be diagonalized by rational coefficients.
The normalization of the two resulting leading singularities then introduces the pair of nested square roots
$\sqrt{q+\sqrt r}$ and $\sqrt{q-\sqrt r}$.

The rational integrals $\vb{I}_{\mathrm R,1}$ and
$\vb{I}_{\mathrm R,2}$ satisfy the coupled LO system
\begin{equation}
\renewcommand{\arraystretch}{1.6}
\label{eq:de-rational-nested-sqrt}
  \dd
  \begin{pmatrix}
    \vb{I}_{\mathrm R,1}\\
    \vb{I}_{\mathrm R,2}
  \end{pmatrix}
  =
  \frac{1}{4}
  \begin{pmatrix}
    \dd\log(q^2-r)
    &
    \omega_1
    \\
    \displaystyle\frac{\omega_1}{r}
    &
    \dd\log(q^2-r)-2\dd\log r
  \end{pmatrix}
  \begin{pmatrix}
    \vb{I}_{\mathrm R,1}\\
    \vb{I}_{\mathrm R,2}
  \end{pmatrix} ,
  \qquad
  \omega_1
  =
  \frac{q\,\dd r-2r\,\dd q}{q^2-r}.
\end{equation}
The fundamental matrix and its inverse are
\begin{equation}
\label{eq:fund-matrix-nested-sqrt}
  \Phi(\vb*{s})
  =
  \begin{pmatrix}
    \sqrt{q+\sqrt r}
    &
    \sqrt{q-\sqrt r}
    \\
    \displaystyle\frac{\sqrt{q+\sqrt r}}{\sqrt r}
    &
    \displaystyle-\frac{\sqrt{q-\sqrt r}}{\sqrt r}
  \end{pmatrix},
  \qquad
  \Phi^{-1}(\vb*{s})
  =
  \begin{pmatrix}
    \displaystyle\frac{1}{2\sqrt{q+\sqrt r}}
    &
    \displaystyle\frac{\sqrt r}{2\sqrt{q+\sqrt r}}
    \\
    \displaystyle\frac{1}{2\sqrt{q-\sqrt r}}
    &
    \displaystyle-\frac{\sqrt r}{2\sqrt{q-\sqrt r}}
  \end{pmatrix}.
\end{equation}
Direct differentiation verifies that $\Phi$ satisfies the LO system in
\cref{eq:de-rational-nested-sqrt}. The two rows of $\Phi^{-1}$ implement
precisely the diagonalization and unit normalization described above.

The LO system in \cref{eq:de-rational-nested-sqrt} has two singular surfaces,
$q^2-r=0$ and $r=0$. Let $\gamma_1$ and $\gamma_2$ be closed loops encircling these two surfaces, respectively.
Explicitly transporting the DE along these loops and comparing the continued fundamental matrix with its value at the base point gives
\begin{equation}
\label{eq:monodromy-nested-sqrt}
  U_{\gamma_1}\Phi
  =
  \Phi
  \begin{pmatrix}
    -1 & 0\\
    0 & -1
  \end{pmatrix},
  \qquad
  U_{\gamma_2}\Phi
  =
  \Phi
  \begin{pmatrix}
    0 & 1\\
    1 & 0
  \end{pmatrix}.
\end{equation}
Comparing these monodromy transformations with the explicit form of $\Phi$,
one observes that the first corresponds to a simultaneous sign flip of the
two outer square roots, whereas the second corresponds to
$\sqrt r\to-\sqrt r$ and exchanges $\sqrt{q+\sqrt r}$ with
$\sqrt{q-\sqrt r}$. The latter is the characteristic doublet action
associated with nested square
roots~\cite{Becchetti:2025oyb,Becchetti:2025qlu}.

The LO homogeneous block is removed by transforming with the inverse
fundamental matrix,
\begin{equation}
\label{eq:Tsqrt-nested}
  \begin{pmatrix}
    \vb{I}_{\mathrm A,1}(\vb*{s},\epsilon)\\
    \vb{I}_{\mathrm A,2}(\vb*{s},\epsilon)
  \end{pmatrix}
  =
  \Phi^{-1}(\vb*{s})
  \begin{pmatrix}
    \vb{I}_{\mathrm R,1}(\vb*{s},\epsilon)\\
    \vb{I}_{\mathrm R,2}(\vb*{s},\epsilon)
  \end{pmatrix}.
\end{equation}
The canonical integrals
$(\vb{I}_{\mathrm A,1},\vb{I}_{\mathrm A,2})$ consequently transform as a
doublet under the monodromy generated by
\cref{eq:monodromy-nested-sqrt}.

A basis $\widetilde f_i(\vb*{s})$ of $\mathbb{Q}$-linear combinations of
iterated integrals can then be constructed as in the square-root case,
following the approach of \cite{Chicherin:2021dyp}. These iterated integrals
have logarithmic kernels involving the nested square roots\footnote{
  Although this may be difficult to verify in practice \cite{Becchetti:2025oyb,FebresCordero:2023pww}.
} and transform in
the same doublet representation as the canonical integrals. The
monodromy-invariant functions are obtained by undoing the algebraic
transformation,
\begin{equation}
  \begin{pmatrix}
    f_1(\vb*{s})\\
    f_2(\vb*{s})
  \end{pmatrix}
  =
  \Phi(\vb*{s})
  \begin{pmatrix}
    \widetilde f_1(\vb*{s})\\
    \widetilde f_2(\vb*{s})
  \end{pmatrix},
\end{equation}
or explicitly,
\begin{align}
  f_1(\vb*{s})
  &=
  \sqrt{q+\sqrt r}\,
  \widetilde f_1(\vb*{s})
  +
  \sqrt{q-\sqrt r}\,
  \widetilde f_2(\vb*{s}),
  \\
  f_2(\vb*{s})
  &=
  \frac{\sqrt{q+\sqrt r}}{\sqrt r}\,
  \widetilde f_1(\vb*{s})
  -
  \frac{\sqrt{q-\sqrt r}}{\sqrt r}\,
  \widetilde f_2(\vb*{s}).
\end{align}
Here, in a slight abuse of notation, the functions $\tilde{f}_i$ are to be understood as $\mathbb{Q}$-linear combinations of iterated integrals that originate from a particular $\epsilon$ order of $\vb{I}_{\mathrm A,i}$. 
Indeed, under $\Phi\to\Phi\mathcal M$, the iterated-integral doublet
transforms as
$\widetilde{\vb f}\to\mathcal M^{-1}\widetilde{\vb f}$, leaving
$\vb f=\Phi\widetilde{\vb f}$ invariant. Since $\Phi$ is independent of
$\epsilon$, the grading by iterated-integral length is preserved.
Moreover, the transformation preserves linear independence over the corresponding algebraic function field.

Before concluding the discussion of the algebraic case, it is worth noting that the square roots encountered above can sometimes be rationalized by a suitable change of variables~\cite{Besier:2019kco}. 
This can be useful for expressing the relevant linear combinations of iterated integrals with algebraic $\dd\log$ kernels in terms of multiple polylogarithms.
In the rationalizing variables, the corresponding DE becomes rational, so the algebraic monodromy is no longer explicit. It has not, however, disappeared.
A rationalization parametrizes the algebraic cover by a rational map from the original variables to the new kinematic variables. Since the cover is nontrivial, the inverse map is generally multivalued.
The original algebraic monodromy is therefore encoded in the different branches of this inverse map, or equivalently in the corresponding transformations of the rationalizing variables.
A function that is monodromy invariant in the original variables must also be invariant under these branch transformations.
Rationalizing the square roots therefore does not help as far as separating the rational and multivalued parts of a scattering amplitude is concerned.

Let us note that similarly to how the nested square roots appear as algebraic solutions of homogeneous leading-order DEs, one may also anticipate that radicals of higher degree or, more generally, solutions of higher degree polynomials may appear.
Their monodromy groups can be constructed similarly as discussed in this section.

\subsection{Elliptic case}
\label{sec:elliptic-case}

The LO homogeneous equations satisfied by Feynman integrals can also have non-algebraic solutions, most notably periods expressed in terms of complete elliptic integrals. 
To handle such a situation, most approaches to elliptic Feynman integrals have focused on constructing an $\epsilon$-factorized basis $\IefactE$ with constant intersection matrix \cite{Adams:2017tga,Dlapa:2022wdu,Gorges:2023zgv,Chen:2025hzq,Yang:2025ofz,Duhr:2025xyy,e-collaboration:2025frv,Bree:2025tug,Bree:2026qww,Duhr:2025lbz}.
The property most important for the present discussion is that
the iterated integrals generated by $\epsilon$ of such bases expansion are expected to be linearly independent over the relevant function
field~\cite{Duhr:2024xsy}. Such bases are commonly referred to as canonical.

A common structural feature of these constructions is that
\begin{equation}
\label{eq:de-elliptic-efact}
  \IefactE(\vb*{s},\epsilon)
  =
  T\big(
    \vb*{s},
    \epsilon,
    \vb*{E}(\vb*{s})
  \big)\,
  \vb{I}_{\mathrm R}(\vb*{s},\epsilon),
  \qquad
  \dd\IefactE(\vb*{s},\epsilon)
  =
  \epsilon\,
  M\big(
    \vb*{s},
    \widetilde{\vb*{E}}(\vb*{s})
  \big)\,
  \IefactE(\vb*{s},\epsilon)  \,,
\end{equation}
where the transformation depends both on $\epsilon$ and on a collection of
transcendental functions $\vb*{E}(\vb*{s})$ that can be expressed through
complete elliptic integrals of the first, second, and third kinds. A general
observation for canonical elliptic bases is that complete elliptic integrals
of the second kind may enter the transformation but are absent from the
$\epsilon$-factorized connection. Thus,
$\widetilde{\vb*{E}}(\vb*{s})$ is obtained from
$\vb*{E}(\vb*{s})$ by removing the complete elliptic integrals of the second
kind. The explicit form of
$T(\vb*{s},\epsilon,\vb*{E}(\vb*{s}))$ can be rather complicated, depends on
the chosen construction, and is currently determined largely case by case. Even identifying the complete set of
transcendental functions $\vb*{E}(\vb*{s})$ required by a given family is
non-trivial~\cite{Duhr:2025xyy,e-collaboration:2025frv,Bree:2026qww}.

In the algebraic case, the rational basis $\vb{I}_{\mathrm R}$ provided a
direct route to monodromy-invariant transcendental functions that separate
the rational and multivalued parts of a scattering amplitude. In the
elliptic case, no comparably simple and universal relation between a rational
basis and an $\epsilon$-factorized basis is currently known. This leaves the
basic question of which rational basis should be used. Equivalently, given
the iterated integrals obtained from $\IefactE$, it is necessary to determine
which monodromy-invariant combinations span precisely the function space
required in \cref{eq:ampl-basis}.
Several prescriptions lead to rational bases with different polynomial or
Laurent-polynomial dependence on $\epsilon$. The geometry-inspired ordering
relation of \cite{e-collaboration:2025frv,Bree:2025tug}, for example,
produces diagonal elliptic blocks of the schematic form
$\epsilon^{-1}M^{(-1)}+M^{(0)}+\epsilon M^{(1)}$. A prescription based on
derivatives of master integrals (see e.g.~\cite{Gorges:2023zgv}),
generally leads instead to a homogeneous block of the form
$M^{(0)}+\epsilon M^{(1)}+\epsilon^2M^{(2)}$.

In this work, we focus on the proposal of
Refs.~\cite{Chaubey:2025adn,Chen:2025hzq}, according to which a rational basis
$\vb{J}_{\mathrm R}$ can be constructed such that the diagonal elliptic block
of its DE is linear in $\epsilon$:
\begin{equation}
\label{eq:de-rational-elliptic}
  \dd\vb{J}_{\mathrm R}(\vb*{s},\epsilon)
  =
  \left[
    M_{\mathrm R}^{(0)}(\vb*{s})
    +
    \epsilon\,M_{\mathrm R}^{(1)}(\vb*{s})
  \right]
  \vb{J}_{\mathrm R}(\vb*{s},\epsilon).
\end{equation}
A systematic integrand-level construction of such a basis is reviewed in the
following sections. Restricting to the subsystem that couples to the elliptic
curve at LO, and suppressing algebraic rows that decouple from it, the
universal LO block takes the schematic form
\begin{equation}
\label{eq:de-rational-elliptic-0}
  \renewcommand{\arraystretch}{2.0}
  M_{\mathrm R}^{(0)}(\vb*{s})
  =
  \frac{1}{2}
  \begin{pmatrix}
    \displaystyle
    \dd\log\!\left(\frac{1}{k}\right)
    &
    \displaystyle
    \dd\log\!\left(\frac{k}{1-k}\right)
    &
    0
    \\
    \displaystyle
    \dd\log\!\left(\frac{1}{k}\right)
    &
    \displaystyle
    \dd\log k
    &
    0
    \\
    \displaystyle
    \dd\log\!\left(\frac{1-m k}{m k}\right)
    &
    \displaystyle
    \frac{1}{1-m k}\,
    \dd\log\!\left(\frac{k(1-m)}{1-k}\right)
    &
    \displaystyle
    \dd\log\!\left(
      \frac{m}{(1-m k)(1-m)}
    \right)
  \end{pmatrix},
\end{equation}
where $k\equiv k(\vb*{s})$ is the parameter of the elliptic curve determined by its roots and $m\equiv m(\vb*{s})$ specifies a marked point. 
In the simplest case, only the upper-left
two-by-two block is present. Each additional marked point contributes
another row of the form of the third row, with its corresponding value of
$m(\vb*{s})$. The diagonal entry in such a row can be removed by an
additional square-root normalization.

A fundamental matrix of \cref{eq:de-rational-elliptic} with $\epsilon=0$  and $M_{\mathrm R}^{(0)}(\vb*{s})$ in \cref{eq:de-rational-elliptic-0} can be chosen as
\begin{equation}
\label{eq:fund-matrix-elliptic}
  \renewcommand{\arraystretch}{2.0}
  \Phi(\vb*{s})
  =
  \begin{pmatrix}
    \psi_1 & \psi_2 & 0
    \\
    \phi_1 & \phi_2 & 0
    \\
    \pi_1^m & \pi_2^m & \sqrt{P(m)}
  \end{pmatrix}
  =
  \begin{pmatrix}
    \mathrm{K}(k)
    &
    \ii\,\mathrm{K}(1-k)
    &
    0
    \\
    \mathrm{E}(k)
    &
    \ii\big[
      \mathrm{K}(1-k)-\mathrm{E}(1-k)
    \big]
    &
    0
    \\
    \Pi(m k,k)
    &
    \displaystyle
    \ii\left[
      \mathrm{K}(1-k)
      +
      \frac{m k}{1-m k}\,
      \Pi\left(
        \frac{1-k}{1-m k},
        1-k
      \right)
    \right]
    &
    \displaystyle
    \frac{\sqrt m}
         {\sqrt{m-1}\sqrt{1-m k}}
  \end{pmatrix}.
\end{equation}
The complete elliptic integrals of the first, second, and third kinds are
defined in \cref{sec:complete-eli}. The first two rows contain the periods and
quasi-periods of the elliptic curve, while the third row contains the periods
of a differential of the third kind associated with the marked point.
Note that, in concrete applications, obtaining a strictly rational basis
$\vb{J}_{\mathrm R}$ may involve a fundamental matrix that includes additional algebraic transformations of \cref{eq:fund-matrix-elliptic} and the corresponding change in the DE in \cref{eq:de-rational-elliptic-0},
which we discuss in more detail in \cref{sec:recap-method}.
This does not substantially affect the subsequent discussion in this section and will be left implicit.

Let us now discuss the monodromy of the fundamental matrix, beginning with its upper-left two-by-two block.
Its connection has singular surfaces at $k=0$ and $k=1$. Explicitly transporting
the LO DE around small loops encircling these surfaces and comparing the
continued fundamental matrix with its value at the base point gives
\begin{equation}
\label{eq:monodromy-elliptic-1}
  U_{\gammacircle{k=0}}\Phi
  =
  \Phi
  \begin{pmatrix}
    1 & 2 & 0\\
    0 & 1 & 0\\
    0 & 0 & 1
  \end{pmatrix},
  \qquad
  U_{\gammacircle{k=1}}\Phi
  =
  \Phi
  \begin{pmatrix}
    1 & 0 & 0\\
    -2 & 1 & 0\\
    0 & 0 & 1
  \end{pmatrix}.
\end{equation}
Thus, analytic continuation mixes the first two columns of $\Phi$ by
integer linear transformations. In contrast to the finite monodromy
representation encountered for nested square roots, the group generated by
the two matrices in \cref{eq:monodromy-elliptic-1} is infinite.
The marked point introduces additional singular surfaces at
$m=0$, $m=1$, and $m=1/k$. Explicit transport around a loop encircling one
of these surfaces gives, depending on the cycle involved,
\begin{equation}
\label{eq:monodromy-elliptic-2}
  U_{\gammacircle{m}}\Phi
  =
  \Phi
  \begin{pmatrix}
    1 & 0 & 0\\
    0 & 1 & 0\\
    -\pi & 0 & -1
  \end{pmatrix}
  \qquad\text{or}\qquad
  U_{\gammacircle{m}}\Phi
  =
  \Phi
  \begin{pmatrix}
    1 & 0 & 0\\
    0 & 1 & 0\\
    0 & -\pi & -1
  \end{pmatrix}.
\end{equation}
Besides changing the sign of the algebraic solution $\sqrt{P(m)}$, these
transformations shift one of the third-kind periods by a multiple of that
solution. For example, the first transformation gives
\begin{equation}
  \Pi(m k,k)
  \longrightarrow
  \Pi(m k,k)
  -
  \frac{\pi\sqrt m}
       {\sqrt{m-1}\sqrt{1-m k}}.
\end{equation}

Inspired by the algebraic case, the LO homogeneous part
$M^{(0)}_\mathrm{R}(\vb*{s})$ in \cref{eq:de-rational-elliptic} can be
removed using the inverse fundamental matrix,
\begin{equation}
  \vb{I}_{\mathcal{E}^\prime}
  (\vb*{s},\epsilon,\vb*{E})
  =
  \Phi^{-1}(\vb*{s},\vb*{E})\,
  \vb{J}_\mathrm{R}(\vb*{s},\epsilon).
\end{equation}
At least within the diagonal DE block, this produces an
$\epsilon$-factorized basis and normalizes the corresponding period matrix
to unity~\cite{Bosma:2017ens,Primo:2017ipr,Frellesvig:2021hkr}. Its solution can therefore be written
order-by-order in $\epsilon$ in terms of iterated integrals.
Under analytic continuation, $\Phi\to\Phi\mathcal{M}$ and hence
$\Phi^{-1}\to\mathcal{M}^{-1}\Phi^{-1}$, where $\mathcal{M}$ denotes any
of the monodromy matrices in
\cref{eq:monodromy-elliptic-1,eq:monodromy-elliptic-2}. Consequently,
$\vb{I}_{\mathcal{E}^\prime}$ transforms in the inverse representation of
the monodromy group. As in the algebraic case, the independence of $\Phi$
from $\epsilon$ implies that this transformation property is inherited
order-by-order by the iterated-integral basis
$\tilde{f}(\vb*{s})$. Monodromy-invariant functions are then
reconstructed as
\begin{align}
  f_1(\vb*{s})
  &=
  \psi_1\,\tilde{f}_1(\vb*{s})
  +
  \psi_2\,\tilde{f}_2(\vb*{s}),
  \\
  f_2(\vb*{s})
  &=
  \phi_1\,\tilde{f}_1(\vb*{s})
  +
  \phi_2\,\tilde{f}_2(\vb*{s}),
  \\
  f_3(\vb*{s})
  &=
  \pi_1^m\,\tilde{f}_1(\vb*{s})
  +
  \pi_2^m\,\tilde{f}_2(\vb*{s})
  +
  \sqrt{P(m)}\,\tilde{f}_3(\vb*{s}).
\end{align}
Here the functions $\tilde{f}_i$ are to be understood as $\mathbb{Q}$-linear combinations of iterated integrals that originate from a particular $\epsilon$ order of $\vb{I}_{\mathcal{E}^\prime}$. 
However, the $\epsilon$-factorized basis obtained in this way does not
possess all the properties desired for the construction of a transcendental
function basis. At each relevant singular point, one of the periods
$\psi_i$ diverges logarithmically. The kernels of the transformed DE
therefore contain logarithmic singularities and are not Fuchsian in the
original kinematic variables. Moreover, when the elliptic curve degenerates, the resulting solutions need
not reduce to $\mathbb{Q}$-linear combinations of logarithmic iterated
integrals~\cite{Frellesvig:2023iwr}. Finally, the non-constant intersection
matrix implies non-trivial linear relations among the iterated
integrals~\cite{Duhr:2026hcs}. These features limit the usefulness of this
integral basis for constructing the independent function basis required in
\cref{eq:ampl-basis}.

A crucial step toward a basis that generates independent iterated integrals
from kernels with at most simple poles, at least locally, is to uncouple
\cref{eq:de-rational-elliptic-0} in a more restricted
way~\cite{Brown:2015ylf,Broedel:2018qkq,Gorges:2023zgv}. Rather than removing the complete
LO connection, one first removes only its semisimple part, leaving a
nilpotent connection. Importantly, this transformation involves only a
single linear combination of the columns of the fundamental matrix,
\begin{equation}
  \vb*{E}
  =
  \begin{pmatrix}
    \psi_c\\
    \phi_c\\
    \pi_c^m
  \end{pmatrix}
  =
  \Phi\,c,
\end{equation}
where $c$ is a constant vector chosen such that $\vb*{E}$ is holomorphic in
a neighborhood of a given singular point~\cite{Frellesvig:2023iwr}. This
choice avoids introducing the non-meromorphic behavior encountered when the
complete inverse fundamental matrix is used.
Without loss of generality, the holomorphic combination may be taken to be
the first column of $\Phi$, corresponding to
$c=(1,0,0)^{\mathrm{T}}$. The transformation is then given by
\begin{equation}
\label{eq:TC}
  \renewcommand{\arraystretch}{1.5}
  T_{\mathcal{C}}(\vb*{E})
  =
  \begin{pmatrix}
    \dfrac{1}{\psi_1} & 0 & 0\\
    -\phi_1 & \psi_1 & 0\\
    -\dfrac{\pi_1^m}{\psi_1\sqrt{P(m)}} & 0
      & \dfrac{1}{\sqrt{P(m)}}
  \end{pmatrix},
\end{equation}
As observed in \cite{Chaubey:2025adn}, the basis
\begin{equation}
  \JomnC(\vb*{s},\epsilon,\vb*{E})
  =
  T_{\mathcal{C}}(\vb*{E})\,
  \vb{J}_\mathrm{R}(\vb*{s},\epsilon)
\end{equation}
satisfies a DE of the form
\begin{equation}
\label{eq:de-elliptic-C}
  \dd\JomnC
  =
  \left[
    (1-n\,\epsilon)\,M^{(0)}_{\mathcal{C}}
    +
    \epsilon\,M^{(1)}_{\mathcal{C}}(\vb*{s},\vb*{E})
  \right]
  \JomnC,
  \qquad
  M^{(0)}_{\mathcal{C}}
  =
  \begin{pmatrix}
    0 & \dd\left(\dfrac{\psi_2}{\psi_1}\right) & 0\\
    0 & 0 & 0\\
    0 &
    \dd\left(
      \pi_2^m-\pi_1^m\dfrac{\psi_2}{\psi_1}
    \right) & 0
  \end{pmatrix},
\end{equation}
where $n$ is an integer. The two matrices have disjoint support:
$(M^{(0)}_{\mathcal{C}})_{ij}\neq0$ implies
$(M^{(1)}_{\mathcal{C}})_{ij}=0$. Thus, every non-vanishing entry has its
$\epsilon$ dependence factorized either as $\epsilon$ or as
$1-n\,\epsilon$. We discuss how to predict the number $n$ in \cref{sec:n-meaning}.

Although this DE is not $\epsilon$-factorized, its LO connection
$M^{(0)}_{\mathcal{C}}$ is nilpotent, with $(M^{(0)}_{\mathcal C})^2=0$. 
Consequently, arbitrarily long sequences of LO kernels cannot occur in the path-ordered
exponential: at order $\epsilon^k$, its expansion contains only finitely
many iterated integrals. The solution can therefore still be constructed
order-by-order in terms of iterated integrals. An explicit example is
presented in \cref{sec:iint-sol}.

An important consequence of using only one holomorphic linear combination of
periods in \cref{eq:TC}, rather than the complete period matrix, is that the
components of $\JomnC$ do not close under monodromy and therefore \emph{do not furnish a representation of the monodromy group}. 
In contrast to the basis $\vb{I}_{\mathcal{E}^\prime}$ discussed above, analytic continuation cannot
be represented by constant matrices acting within $\JomnC$. This is
conceptually different from the algebraic case and is closely related to the
behavior of elliptic Feynman integrals under modular transformations, where a
change of periods must generally be accompanied by a corresponding
redefinition of the master integrals~\cite{Weinzierl:2020fyx}.
Nevertheless, the simple and universal form of $T_{\mathcal{C}}$ allows
monodromy-invariant functions to be reconstructed by applying
$T_{\mathcal{C}}^{-1}$ to the iterated-integral basis at each order in $\epsilon$:
\begin{equation}
  \label{eq-monodromy-inv-elliptic}
  \begin{aligned}
    f_1(\vb*{s})
    &=
    \psi_1\,\tilde{f}_1(\vb*{s}),
    \\
    f_2(\vb*{s})
    &=
    \phi_1\,\tilde{f}_1(\vb*{s})
    +
    \frac{1}{\psi_1}\,\tilde{f}_2(\vb*{s}),
    \\
    f_3(\vb*{s})
    &=
    \pi_1^m\,\tilde{f}_1(\vb*{s})
    +
    \sqrt{P(m)}\,\tilde{f}_3(\vb*{s}).
  \end{aligned}
\end{equation}
Here the functions $\tilde{f}_i$ are to be understood as $\mathbb{Q}$-linear combinations of iterated integrals that originate from a particular $\epsilon$ order of $\vb{J}_{\mathcal{C},i}$. 
Since $T_{\mathcal{C}}$ is independent of $\epsilon$,
this reconstruction applies separately at every order in $\epsilon$, as in
the algebraic case.

This construction achieves the desired separation between the rational and
multivalued parts of the amplitude. Two major problems nevertheless remain.
First, the discussion so far has concerned only the diagonal DE blocks. The
extension to the off-diagonal blocks is one of the main results of this paper
and is postponed to \cref{sec:inh-de}. Second,
$M_{\mathcal{C}}^{(1)}(\vb*{s},\vb*{E})$ depends on the quasi-period
$\phi_1$, which can be expressed algebraically in terms of $\psi_1$ and its derivatives.
The kernels in \cref{eq:de-elliptic-C} therefore satisfy relations involving
total derivatives, and the resulting iterated integrals need not be linearly
independent over the relevant function field. Constructing the basis
$\tilde{f}_i(\vb*{s})$ consequently requires systematically accounting for
integration-by-parts relations, similar to those considered in
\cite{Badger:2021owl}, as well as identities between iterated integrals with
non-constant coefficients.

An alternative is to perform a sequence of further transformations
$T(\vb*{s},\epsilon,\vb*{E})$ leading to an $\epsilon$-factorized DE with a
constant intersection matrix, which is expected to generate independent
iterated integrals~\cite{Duhr:2024xsy,Duhr:2025xyy}. Following the
observations of \cite{Chen:2025hzq,Yang:2025ofz}, the first step starting
from \cref{eq:de-elliptic-C} is to rescale the second component of $\JomnC$,
associated with the complete elliptic integral of the second kind in
\cref{eq:fund-matrix-elliptic}, by
$(1-n\epsilon)/\epsilon$. This brings the DE into Laurent-polynomial form in
$\epsilon$. The remaining non-$\epsilon$-factorized entries can then, in
principle, be removed systematically by solving a sequence of auxiliary
inhomogeneous DEs (see e.g.~\cite{e-collaboration:2025frv,Bree:2025tug,Gorges:2023zgv}). 
Related constructions at the
integrand level were developed in
\cite{Yang:2025ofz,Forner:2026vby}.
For the present discussion, the important point is that the resulting
transformation $T(\vb*{s},\epsilon,\vb*{E})$ necessarily depends on both
$\epsilon$ and the functions $\vb*{E}$ entering \cref{eq:TC}. Moreover, the
set $\vb*{E}$ may have to be enlarged by additional functions with
non-trivial monodromy~\cite{e-collaboration:2025frv,Duhr:2025lbz,
Becchetti:2025oyb,Duhr:2025xyy}. These functions satisfy inhomogeneous DEs
analogous to the one encoded by the third row of
\cref{eq:de-rational-elliptic-0}, after the corresponding square-root
normalization. Such extensions arise in particular when
$\epsilon$-factorizing the off-diagonal DE blocks.
A common approach to this situation is to enlarge the field of coefficients
in \cref{eq:ampl-basis} by adjoining all functions in $\vb*{E}$, while
taking the basis functions to be $\mathbb{Q}$-linear combinations of the
iterated-integral solutions of \cref{eq:de-elliptic-efact}. Since the
functions in $\vb*{E}$ have non-trivial monodromy, however, the resulting
coefficients are themselves multivalued. This is incompatible with the
separation between rational and multivalued contributions required here.

Taking these observations together, we propose the rational basis
$\vb{J}_\mathrm{R}$ constructed following
\cite{Chaubey:2025adn,Chen:2025hzq} as a natural rational-basis candidate
for elliptic Feynman integrals. It satisfies the homogeneous DE
in \cref{eq:de-rational-elliptic}; applying $T_{\mathcal{C}}$ produces the
nilpotent DE \cref{eq:de-elliptic-C}, from which the
monodromy-invariant functions in
\cref{eq-monodromy-inv-elliptic} can be reconstructed. Relations among the
functions $\tilde{f}_i(\vb*{s})$ may then be resolved by expressing them in
terms of iterated integrals obtained from the basis $\IefactE$.
As in the nested-square-root case, the resulting monodromy-invariant
transcendental basis is not, in general, given by $\mathbb{Q}$-linear
combinations of the iterated integrals obtained directly from the
$\epsilon$-factorized DE. Furthermore, because the transformation
$T(\vb*{s},\epsilon,\vb*{E})$ depends explicitly on $\epsilon$, it can mix
iterated integrals of different lengths between different orders in
$\epsilon$. There is therefore no obvious grading of the
monodromy-invariant functions by iterated-integral length.

\begin{table}[ht]
  \newcommand{\picwidth}{12ex}
  \newcommand{\centeredfig}[1]{$\vcenter{\hbox{\includegraphics[width=\picwidth]{#1}}}$}
  \renewcommand{\arraystretch}{2.1}
  \centering
  \begin{tabular}{C{18ex}  C{20ex}   C{20ex} C{12ex}}
    \toprule
    Graph  & Number of MIs on the elliptic sector   & Number of MIs in subsectors & Acronym \\
    \midrule
    \centeredfig{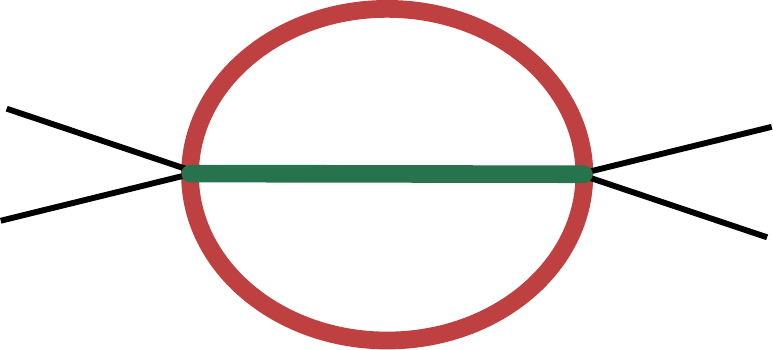}         & 3    & 2 & \texttt{SR} \\[1ex]
    $\vcenter{\hbox{ $\begin{multlined}  \includegraphics[height=6ex]{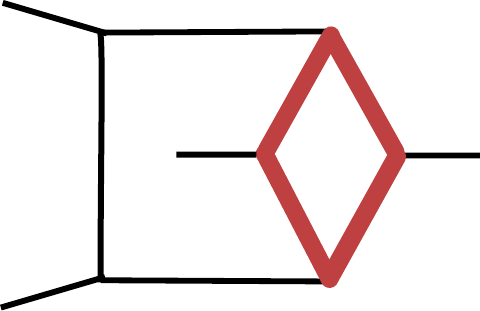} \\[-3ex] \;\includegraphics[height=5.5ex]{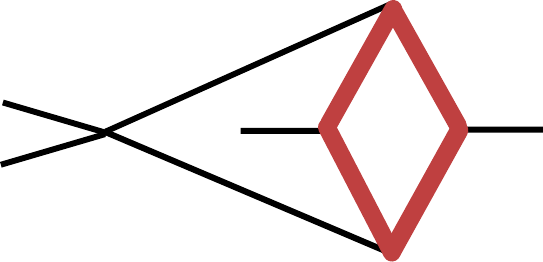} \end{multlined}$  }}$   & 4+2  & 30 & \texttt{D}    \\
    \centeredfig{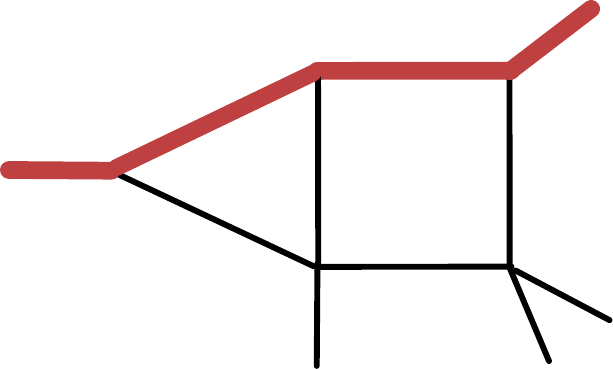}              & 3    & 25 & \texttt{TT}   \\
    \bottomrule
  \end{tabular}
  \caption{The examples of Feynman integral families considered in this work. Thick lines denote massive lines (red, green correspond to different masses)}
  \label{tab:examples-summary}
\end{table}

\begin{example}
\label{ex:functions}

We illustrate the interplay between rationality and monodromy invariance using the two-mass sunrise family, denoted by \texttt{SR} in \cref{tab:examples-summary},
which is defined by
\begin{equation}
  \vb*{\rho}_{\texttt{SR}} = \left\{ \ell_1^2-m^2,\ \ell_2^2-M^2,\ (\ell_1-\ell_2-p)^2-m^2,\ \ell_1\cdot p,\ \ell_2\cdot p \right\}  \,,
\end{equation}
and we set $p^2 = 1$.
We begin with the rational basis $\vb{J}_\mathrm{R}$ (we explain how it is constructed in \cref{ex:numerator-ansatz}):
\begin{equation}
\label{eq:rational_basis_sunrise2m}
  \vb{J}_\mathrm{R}  = \begin{pmatrix}
      G^{0, 1, 1, 0, 0}_{\texttt{SR}} \\
      G^{1, 0, 1, 0, 0}_{\texttt{SR}} \\
      G^{1,1,1,0,0}_{\texttt{SR}}  \\
      2(m^2\,M) \, G^{1,1,1,0,0}_{\texttt{SR}} + (2 m^2 - M^2-1) \, G^{1,1,1,0,-1}_{\texttt{SR}} -2\, G^{1,1,1,0,-2}_{\texttt{SR}} +   G^{1,0,1,0,0}_{\texttt{SR}}   \\
      G^{1, 1, 1, 0, -1}_{\texttt{SR}} \\
    \end{pmatrix}, 
\end{equation}
The rational DE is given by 
\begin{equation}
  \dd \vb{J}_\mathrm{R} =  \begin{pmatrix}
\epsilon\,\omega_{1}
& 0
& 0
& 0
& 0
\\[2mm]

0
& \epsilon\,\omega_{2}
& 0
& 0
& 0
\\[2mm]

\epsilon\,\omega_{3}
& \epsilon\,\left(-\omega_{3}+\omega_{5}\right)
& \omega_{4} + \epsilon\,\alpha
& (3\epsilon-1)\,\omega_{5}
& \epsilon\,\omega_{6}
\\[2mm]

\epsilon\,\omega_{7}
& \epsilon\,\left(-\omega_{4}-\omega_{7}\right)
& \omega_{8}-\epsilon\,\frac{\omega_{9}}{3}
& (3\epsilon-1)\,\omega_{4}
& \epsilon\,\omega_{10}
\\[2mm]

\epsilon\,\omega_{11}
& \epsilon\,\left(-\omega_{11}-\frac{\omega_{3}}{2}\right)
& \frac{\omega_{7}}{8}-\epsilon\,\frac{\omega_{12}}{3}
& (3\epsilon-1)\,\frac{\omega_{3}}{2}
& \epsilon\,\omega_{13}
  \end{pmatrix} \; \vb{J}_\mathrm{R}  \,,
\end{equation}
where $\omega_i$ are rational one-forms whose explicit form is given in \cref{sec:rat-one-form-aux}, and
\begin{equation}
  \alpha =  \frac{1}{3} \qty(5\omega_{1}-5\omega_{2}-10\omega_{3}-9\omega_{4} +6\omega_{5}-4\omega_{6} +5\omega_{11}+5\omega_{13})  \,.
\end{equation}
An $\epsilon$-factorized DE is obtained through the transformation
\begin{equation}
\renewcommand{\arraystretch}{1.1}
\setlength{\arraycolsep}{8pt}
\label{eq:rational-to-canonical}
  \IefactE
  =
  \begin{pmatrix}
    1 & 0 & 0 & 0 & 0
    \\
    0 & 1 & 0 & 0 & 0
    \\
    0 & 0 & \dfrac{1}{\psi} & 0 & 0
    \\[2mm]
    -4\pi^\infty
    &
    2\bigl(\psi+2\pi^\infty\bigr)
    &
    A\psi
      -\dfrac{\phi}{2\epsilon}
      -\dfrac{6(\pi^\infty)^2}{\psi}
    &
    \dfrac{(1-3\epsilon)}{\epsilon} 2 \psi
    &
    4\bigl[(1-m^2+M^2)\psi+3\pi^\infty\bigr]
    \\[2mm]
    0 & 0 & -\dfrac{\pi^\infty}{\psi} & 0 & 1
  \end{pmatrix} \, \vb{J}_{\mathrm{R}}  \,,
\end{equation}
with $A=1+M^4-4m^2\bigl[1+M(M-3)\bigr]$, and is given by
\begin{equation}
  \renewcommand{\arraystretch}{1.4}
  \dd \IefactE  = \epsilon\,\begin{pmatrix} 
 \eta_{1} & 0 & 0 & 0 & 0 \\
 0 & \eta_{2} & 0 & 0 & 0 \\
 \eta_{3} & -\eta_{3} & \eta_{4} & \eta_{5} & -3 \eta_{3} \\
 \eta_{6} & \eta_{7} & \eta_{8} & \eta_{4} & \eta_{9} \\
 \eta_{10} & -\eta_{10} & \frac{\eta_{9}}{12} & -\frac{\eta_{3}}{4} & \frac{3 \eta_{2}}{2}-3 \eta_{10} \\
  \end{pmatrix} \; \IefactE
\end{equation}
with $\vb*{E}=\{\psi,\phi,\pi^\infty\}$ and the 10 $\mathbb{Q}$-linear independent one-forms
\begin{equation}
  \begin{aligned}
    \eta_1 &= \bar{\omega}_{1}, \quad \eta_2 = \bar{\omega}_{2} \,, \quad
    \eta_3 = -4 \bar{\omega}_{8}\,\frac{1}{\psi} - 4 \bar{\omega}_{9}\,\frac{\pi^{\infty}}{\psi^2}\,,\\
    \eta_4 &= \bar{\omega}_{1} + \frac{\bar{\omega}_{4}}{4} - 3 \bar{\omega}_{6} + 12 \bar{\omega}_{8} \frac{\pi^{\infty}}{\psi} + 6 \bar{\omega}_{9} \frac{\left(\pi^{\infty}\right)^2}{\psi^2}\,, \quad
    \eta_5 = -\bar{\omega}_{9}\frac{1}{\psi^2} \,,\\
    \eta_6 &= \bar{\omega}_{3} \psi + \bar{\omega}_{4} \pi^{\infty} + 72 \bar{\omega}_{8} \frac{\left(\pi^{\infty}\right)^2}{\psi} + 24 \bar{\omega}_{9} \frac{\left(\pi^{\infty}\right)^3}{\psi^2} \,,\\
    \eta_7 &= (2 \bar{\omega}_{2} - \bar{\omega}_{3})\psi + (-4 \bar{\omega}_{1} + 4 \bar{\omega}_{2} - \bar{\omega}_{4}) \pi^{\infty} - 72 \bar{\omega}_{8} \frac{\left(\pi^{\infty}\right)^2}{\psi} - 24 \bar{\omega}_{9} \frac{\left(\pi^{\infty}\right)^3}{\psi^2} \,,\\
    \eta_8 &= \bar{\omega}_{5}\,\psi^2 + 24 \bar{\omega}_{7} \,\psi \, \pi^{\infty} + (-12 \bar{\omega}_{1} + 18 \bar{\omega}_{2} - 3 \bar{\omega}_{4})\left(\pi^{\infty}\right)^2 - 144 \bar{\omega}_{8} \frac{\left(\pi^{\infty}\right)^3}{\psi} - 36 \bar{\omega}_{9} \frac{\left(\pi^{\infty}\right)^4}{\psi^2} \,,\\
    \eta_9 &= 12 \bar{\omega}_{7} \,\psi + (-12 \bar{\omega}_{1} + 18 \bar{\omega}_{2} - 3 \bar{\omega}_{4})\pi^{\infty} - 216 \bar{\omega}_{8} \frac{\left(\pi^{\infty}\right)^2}{\psi} - 72 \bar{\omega}_{9} \frac{\left(\pi^{\infty}\right)^3}{\psi^2} \,, \quad
    \eta_{10} = \bar{\omega}_{6} + 8 \bar{\omega}_{8} \frac{\pi^{\infty}}{\psi} + 4 \bar{\omega}_{9} \frac{\left(\pi^{\infty}\right)^2}{\psi^2} \,,
  \end{aligned}
\end{equation}
where only the dependence on $\vb*{E}$ is shown explicitly and the auxiliary rational one-forms $\bar{\omega}_i$ are given in \cref{sec:rat-one-form-aux}.

To mimic the IBP reduction arising in an amplitude computation, we consider the
reduction of the tensor integral $G^{1,1,1,-2,-1}_\texttt{SR} $ and retain its expansion through the
finite order in $\epsilon$. The required terms in the solution are
\begin{equation}
  \renewcommand{\arraystretch}{2}
  \IefactE = 
  \begin{pmatrix}
    1+\epsilon \left(b_1^{(1)}+I_\gamma(\eta_{1})\right)+\epsilon^2 \left(\frac{\left(b_1^{(1)}\right)^2}{2}+I_\gamma(\eta_{1}) b_1^{(1)}+I_\gamma(\eta_{1}, \eta_{1})+\zeta_2\right)+\order{\epsilon^3} \\
 1+\epsilon \left(b_2^{(1)}+I_\gamma(\eta_{2})\right)+\epsilon^2 \left(\frac{\left(b_2^{(1)}\right)^2}{2}+I_\gamma(\eta_{2}) b_2^{(1)}+I_\gamma(\eta_{2}, \eta_{2})+\zeta_2\right)+ \order{\epsilon^3} \\
 \epsilon^2 \left(b_3^{(2)} + (b_1^{(1)}-b_2^{(1)})\,I_\gamma(\eta_{3}) + b_4^{(1)}\,I_\gamma(\eta_{5}) +I_\gamma(\eta_{1} - \eta_2, \eta_{3})+I_\gamma(\eta_{6}+\eta_{7}, \eta_{5})\right)+ \order{\epsilon^3}
   \\
 \epsilon \left(b_4^{(1)}+I_\gamma(\eta_{6})+I_\gamma(\eta_{7})\right)  + \order{\epsilon^2} \\ 
 \epsilon^2 \left(b_5^{(2)} - \frac{b_4^{(1)}}{4}\, I_\gamma(\eta_{3}) + (b_1^{(1)}-b_2^{(1)})\,I_\gamma(\eta_{10})-\frac{1}{4} I_\gamma(\eta_{6} + \eta_{7}, \eta_{3})+I_\gamma(\eta_{1}-\eta_{2}, \eta_{10})\right)+ \order{\epsilon^3} \\
  \end{pmatrix}  \,,
\end{equation}
where $b_i^{(w)}$ are the base values for integral $i$ at $\epsilon$ order $w$.
It is evident that at the required orders  there are no additional linear relations among the $\epsilon$ orders of $\IefactE$ after accounting for the shuffle identities.
Consequently, \cref{eq:linear-dependencies} is readily trivialized by choosing the basis
\begin{equation}
  \begin{aligned}
    \tilde{f}_1 &= {\IefactE}_1^{(1)}  \,,\qquad \tilde{f}_2 = {\IefactE}_2^{(1)}  \,,\qquad \tilde{f}_3 = {\IefactE}_4^{(1)}  \,, \\
    \tilde{f}_4 &= {\IefactE}_1^{(2)} = \frac{\tilde{f}^2_1}{2} + \zeta_2   \,,\qquad \tilde{f}_5 = {\IefactE}_2^{(2)} = \frac{\tilde{f}^2_2}{2} + \zeta_2 \,,   \\
    \tilde{f}_6 &= {\IefactE}_3^{(2)}  \,,\quad \tilde{f}_7 = {\IefactE}_5^{(2)}  \,.  \\
  \end{aligned}
\end{equation}
The result for the reduction of integral $G^{1,1,1,-2,-1}_\texttt{SR}$ can now be written as 
\begin{equation}
  \begin{aligned}
    G^{1,1,1,-2,-1}_\texttt{SR} = &\frac{1}{\epsilon^2}+\frac{1}{\epsilon} \left(\tilde{f}_2-\frac{1}{2}\right)\;+\;\frac{8 m^2-20 M^2-7}{4}  \;+\; \tilde{f}_5
     \;+\;\tilde{f}_1 \left((-2 m^2+5 M^2+1) \frac{\red{\pi^{\infty}}}{\red{\psi}}+ 3 M^2\right) \\
    &+\tilde{f}_2 \left(\frac{-6 M^2-1}{2}+(2 m^2-5 M^2-1) \frac{\red{\pi^{\infty}}}{\red{\psi}}\right)
     \;+\; \red{\tilde{f}_3} \frac{(-2 m^2+5 M^2+1)}{4} \frac{1}{\red{\psi}} \\
    &+\red{\tilde{f}_6} \left( p_1(m,M)\,\red{\psi} - \frac{\left(2 m^2-5 M^2-1\right)}{8} \red{\phi}  + \left(2 m^4+4 m^2+M^4+1\right) \red{\pi^{\infty}} \right) \\
    &+\red{\tilde{f}_7} \left(2 m^4+4 m^2+M^4+1\right) \qquad + \quad \order{\epsilon^1}
  \end{aligned}
\end{equation}
with $p_1(m,M) = M(2 m^4-5 m^2 M^2+2 m^2 M-m^2+M^3+M)$ and we highlighted the functions transformed by the monodromy group given in \cref{eq:monodromy-elliptic-1,eq:monodromy-elliptic-2}.
As we can see, the coefficients of $\tilde{f}_i$ depend on $\vb*{E}$, and such transformations will lead to non-trivial reshuffling of the coefficients in this representation.
If we instead write the result through the rational basis, we obtain 
\begin{equation}
  \begin{aligned}
    G^{1,1,1,-2,-1}_\texttt{SR} = \frac{1}{\epsilon^2}+&\frac{1}{\epsilon} \left(f_2-\frac{1}{2}\right)\;+\;\frac{-4 m^2+10 M^2-1}{4} \;+\; 3 M^2\, f_1 \;-\; \left(m^2+\frac{M^2}{2}\right)\,f_2   \\
    &-\frac{\left(2 m^2-5 M^2-1\right)}{2}\,f_3 \;+\;f_5 \;+\; p_1(m,M)\,f_6 \;+\; \left(2 m^4+4 m^2+M^4+1\right)\,f_7  \quad + \quad \order{\epsilon^1}  \,,
\end{aligned}
\end{equation}
where the coefficients are rational and the functions $f_i$ are by construction monodromy-invariant. They are defined as
\begin{equation}
  \begin{aligned}
    f_1 &= \vb{J}_{\mathrm{R}, 1}^{(1)} =  \tilde{f}_1, \qquad f_2 = \vb{J}_{\mathrm{R}, 2}^{(1)} = \tilde{f}_2,  \qquad f_5 = \vb{J}_{\mathrm{R}, 2}^{(2)} = \tilde{f}_5, \\
    f_6 &= {\vb{J}_\mathrm{R}}_3^{(2)} = \psi \, \tilde{f}_6  \,, \\
    f_3 &= {\vb{J}_\mathrm{R}}_4^{(2)} =  -3 - \tilde{f}_2 \;+\; \frac{\phi}{4}\,\tilde{f}_6 \, +  \frac{1}{2 \psi} \, \tilde{f}_3 \;+\; 2 \frac{\pi^{\infty}}{\psi} (\tilde{f}_{1} - \tilde{f}_2)  \,,  \\
    f_7 &= {\vb{J}_\mathrm{R}}_5^{(2)} = \tilde{f}_7 + \pi^{\infty}\, \tilde{f}_6  \,.
  \end{aligned}
\end{equation}
We see that $f_3$ is a rather non-trivial linear combination of iterated
integrals with coefficients in
$\mathbb{Q}(\vb*{s},\vb*{E})$. Compared with the basis $\tilde f_i$, we
are forced to replace a length-one function by a function involving
iterated integrals of length up to two. The first two terms in $f_3$ are
separately monodromy invariant and can be removed by taking
$\mathbb{Q}$-linear combinations with the other functions, while the
remaining combination of terms with non-trivial monodromy appears to be
minimal. In particular, $f_3$ cannot be replaced, through rational linear
combinations, by a function expressed only in terms of length-one iterated
integrals. However, assuming that the $\tilde f_i$ are independent over the
relevant function field, the functions $f_i$ are also independent.
  
\end{example}

\section{Rational bases from elliptic leading singularities}
\label{sec:recap-method}

We have argued that rational bases play a central role in constructing the monodromy-invariant functions relevant for scattering amplitudes, while retaining information about the underlying geometry of the integral family.
Motivated by this observation, in Ref.~\cite{Chaubey:2025adn} (see also~\cite{Chen:2025hzq}) we proposed a generalization of the $\dd\log$ integrand construction that uses elliptic leading singularities as a guiding principle for selecting candidate rational bases in elliptic sectors. 
That construction was formulated primarily on the maximal cut and therefore determined the diagonal blocks of the differential equations, apart from a few simple examples involving subsector contributions.
In this section, we review the construction in detail. 
Our approach to the off-diagonal DE blocks is discussed in the next section.

Recall that the $\dd\log$ construction proceeds in three steps. First, one constrains the ansatz in \cref{eq:ansatz} so that its integrand is a logarithmic $k$-form of the type shown in \cref{eq:dlog-integrands}. Second, one diagonalizes its leading singularities so that each candidate has only one non-vanishing leading singularity. Third, this leading singularity is normalized to unity. In the nested-square-root case, the first step can still be performed rationally, but the diagonalization in the second step requires algebraic coefficients. In the elliptic case, the obstruction appears already in the first step.

Indeed, when attempting to express the ansatz in \cref{eq:ansatz} as a logarithmic $k$-form, one may find a contribution of the schematic form~\cite{Bourjaily:2017bsb,Broedel:2018qkq,Gorges:2023zgv}
\begin{equation}
\label{eq:elliptic-obstruction}
  \sum_{\vb{i}}
  n_{\vb{i}}(\vb*{s})\,
  G^{\vb{i}}
  \;\xrightarrow{\mathcal{P}}\;
  \frac{f[n](\vb*{s},z)\,\dd z}{\sqrt{P(z)}}
  \wedge
  \left[
    \sum_j
    \bigwedge_{i=1}^{k-1}
    \dlog\beta_{j,i}
  \right]
  +
  \sum_j
  l_j[n](\vb*{s})
  \left[
    \bigwedge_{i=1}^{k}
    \dlog\alpha_{j,i}
  \right],
\end{equation}
where $z=z_k$, the function $f[n](\vb*{s},z)$ is rational in $z$ and depends linearly on the coefficients of the ansatz, and
\begin{equation}
  P(z)
  =
  \prod_{i=1}^{4}
  \left(z-r_i(\vb*{s})\right)
\end{equation}
is a quartic polynomial defining an elliptic curve. The second term in \cref{eq:elliptic-obstruction} is the usual logarithmic part of the ansatz and will be omitted in the remainder of this section.
Using the parametrization in \cref{eq:parametrization}, the dimensionally regulated integrand can be viewed schematically as $u_\epsilon\,\omega$, where all $\epsilon$-independent algebraic factors are included in the algebraic differential $\omega$ and $u_0=1$.
The factor $u_\epsilon$ defines a twisted differential $\nabla_\epsilon=\dd+\dd\log u_\epsilon\wedge$, which reduces to the ordinary exterior derivative at $\epsilon=0$.
Working with the leading-order integrand therefore amounts to first studying the untwisted algebraic cohomology.
This retains the underlying algebraic variety and the pole structure of its differential forms, while the dependence on the dimensional regulator enters through a subsequent deformation to twisted cohomology.
The usual $\dd\log$ construction can be understood in the same way, since the logarithmic forms are identified in the untwisted limit before the dimensional deformation is restored.
We adopt the analogous heuristic for elliptic integrals, assuming that the untwisted algebraic cohomology retains the information needed to select the rational basis candidates relevant to the full dimensionally regulated problem.

The first term cannot, in general, be fully localized on a product of small circles: after localizing the first $k-1$ variables, the remaining integration takes place along non-contractible cycles of the elliptic curve. A basis of such cycles may be chosen, without loss of generality, as $\gamma_1$ and $\gamma_2$, encircling the pairs of roots $(r_2,r_3)$ and $(r_3,r_4)$, respectively. If $f[n](\vb*{s},z)$ has an additional simple pole at $z=a$, one must also consider a small cycle $\gammacircle{a}$ surrounding this pole.

Before proceeding further, let us note that there are special cases in which the elliptic obstruction in~\cref{eq:elliptic-obstruction} disappears. An ordinary logarithmic candidate is obtained if there exists a non-trivial choice of the coefficients for which $f[n](\vb*{s},z)=0$ while at least one logarithmic leading singularity remains nonzero. It may also be possible to express the remaining algebraic one-form as $\dd\log \left( \frac{A(z)-\sqrt{P(z)}} {A(z)+\sqrt{P(z)}} \right)$ with some rational function $A(z)$.
Such candidates remain logarithmic $k$-forms, at least at leading order in $\epsilon$ using the parametrization \eqref{eq:parametrization}. It would be interesting to understand whether they evaluate to iterated integrals with logarithmic kernels, which we leave for future work.

In the following, we assume that the ansatz has been brought to the form of \cref{eq:elliptic-obstruction}. This can be achieved, for example, by applying the iterative procedure of Ref.~\cite{Henn:2020lye} to the first $k-1$ variables and stopping before the final variable $z$ is considered. In practice, obtaining this form can depend sensitively on the parametrization, the ordering of the integration variables, and suitable changes of variables of the type discussed in Ref.~\cite{Henn:2020lye}. For the examples considered here, the standard Baikov parametrization is sufficient and works particularly well after the maximal cut has eliminated most of the integration variables. 
Nevertheless, a decomposition of the form \cref{eq:elliptic-obstruction} need not always exist and we leave the study of such cases for future work.

We now focus on the final variable $z=z_k$ and replace the usual diagonalization of logarithmic leading singularities by matching the remaining algebraic one-form to a judiciously chosen basis 
\begin{equation}
\label{eq:integrand-basis}
  \omega_\kappa
  =
  \frac{N_\kappa(z)\,\dd z}{\sqrt{P(z)}},
  \qquad
  \begin{aligned}
    N_\psi(z)
    &=
    1,
    \\
    N_\phi(z)
    &=
    -2z^2
    +
    (r_1+r_2+r_3+r_4)z
    -
    (r_1r_4+r_2r_3),
    \\
    N_{\pi^\infty}(z)
    &=
    z,
    \\
    N_{\pi^{a}}(z)
    &=
    \frac{1}{z-a}.
  \end{aligned}
\end{equation}
The forms $\omega_\psi$ and $\omega_\phi$ are Abelian differentials of the first and second kind, respectively: $\omega_\psi$ is holomorphic, whereas $\omega_\phi$ has double poles at infinity but vanishing residues. 
Their integrations along the basis cycles give the periods and quasi-periods of the elliptic curve that are closely related to the conventions of Ref.~\cite{Weinzierl:2022eaz}.
The one-forms $\omega_{\pi^\infty}$ and $\omega_{\pi^{a}}$ are differentials of the third kind. The former has simple poles at infinity, with residues $\pm1$, while the latter has simple poles $z=a$, with residues
\begin{equation}
  \mathop{\mathrm{Res}}_{z=a}
  \omega_{\pi^{a}}
  =
  \pm\frac{1}{\sqrt{P(a)}}.
\end{equation}
Relative to the definitions of Ref.~\cite{Chaubey:2025adn}, the normalizations of $\omega_\phi$ and $\omega_{\pi^{a}}$ have been chosen to make the rationality properties of the corresponding integral candidates manifest.\footnote{
  With this choice, $N_\phi(z)$ is rational whenever
  $r_1r_4+r_2r_3\in\mathbb{Q}(\vb*{s})$. This holds, in particular, when
  $P(z)$ factorizes over $\mathbb{Q}(\vb*{s})$ into quadratic polynomials
  whose roots are paired as $(r_1,r_4)$ and $(r_2,r_3)$, as it happens in the
  examples considered here. If $P(z)$ is irreducible, one should instead
  construct a second-kind numerator involving only fully symmetric
  combinations of its roots. We leave a systematic treatment of this case
  for future work.
}

Candidate rational integrals $\vb{J}_{\mathrm{R},\kappa}$ are then obtained by solving for rational coefficients $n_{\vb{i}}^\kappa(\vb*{s})$ such that the algebraic part of the ansatz matches the corresponding numerator in \cref{eq:integrand-basis}, while its logarithmic leading singularities vanish:
\begin{equation}
\label{eq:eLS-candidates}
  \vb{J}_{\mathrm{R},\kappa}
  =
  \sum_{\vb{i}}
  n_{\vb{i}}^\kappa(\vb*{s})\,
  G^{\vb{i}},
  \qquad
  f[n^\kappa](\vb*{s},z)
  =
  N_\kappa(z),
  \qquad
  l_j[n^\kappa](\vb*{s})
  =
  0.
\end{equation}

The elliptic leading singularities are defined by integrating the basis
one-forms in \cref{eq:integrand-basis} over the basis cycles.
In addition
to the two elliptic cycles, we include a small contour around infinity and
one around an arbitrary marked point $a$. The resulting pairings give rise to the period matrix and are given by
\begin{equation}
\label{eq:period-matrix}
  \renewcommand{\arraystretch}{1.4}
  \setlength{\arraycolsep}{8pt}
  \begin{array}{l|cccc}
    \toprule
    \omega_\kappa
    & \int_{\gamma_1}
    & \int_{\gamma_2}
    & \int_{\gammacircle{\infty}}
    & \int_{\gammacircle{a}}
    \\
    \midrule
    \omega_\psi
    & \psi_1
    & \psi_2
    & 0
    & 0
    \\
    \omega_\phi
    & \phi_1
    & \phi_2
    & 0
    & 0
    \\
    \noalign{\vskip 5pt}
    \omega_{\pi^\infty}
    & \pi^\infty_1
    & \pi^\infty_2
    & 1
    & 0
    \\
    \omega_{\pi^a}
    & \pi^a_1
    & \pi^a_2
    & 0
    & \frac{1}{\sqrt{P(a)}}
    \\
    \bottomrule
  \end{array}
\end{equation}
where the small-circle integrals are understood as normalized contour
integrals. 
The location $a$ is determined by the pole structure of $f[n](\vb*{s},\vb*{z})$ in \cref{eq:elliptic-obstruction}.
An explicit representation of the elliptic leading singularities through complete elliptic integrals of three kinds is presented in \cref{sec:complete-eli}.

The periods $\psi_i$ and quasi-periods $\phi_i$ satisfy the Legendre identity
\begin{equation}
\label{eq:wronskian}
  \psi_1\phi_2-\psi_2\phi_1
  =
  2\pi\ii
\end{equation}
and the coupled differential equation
\begin{equation}
\label{eq:psi-phi-de}
  \renewcommand{\arraystretch}{1.6}
  \setlength{\arraycolsep}{6pt}
  \dd
  \begin{pmatrix}
    \psi_i\\
    \phi_i
  \end{pmatrix}
  =
  \frac{1}{2}
  \begin{pmatrix}
    \dd\log\qty(\frac{1}{r_{31}r_{42}\,k})
    &
    \frac{1}{r_{31}r_{42}}\,
    \dd\log\qty(\frac{k}{1-k})
    \\[2mm]
    r_{31}r_{42}\,
    \dd\log\qty(\frac{1}{k})
    &
    \dd\log\qty(r_{31}r_{42}\,k)
  \end{pmatrix}
  \begin{pmatrix}
    \psi_i\\
    \phi_i
  \end{pmatrix},
\end{equation}
where
\begin{equation}
  k
  =
  \frac{r_{32}r_{41}}{r_{31}r_{42}}.
\end{equation}
The third-kind periods satisfy
\begin{align}
\label{eq:pi-de}
  \dd\pi_i^\infty
  &=
  \frac{1}{2}
  \left[
    \sum_{(p,q)\in\{(1,4),(2,3)\}}
    \frac{r_p\,\dd r_q-r_q\,\dd r_p}{r_p-r_q}
  \right]\psi_i
  -
  \frac{1}{2}
  \left[
    \sum_{j=1}^{4}
    \frac{r_j\,\dd r_j}{\Delta_{4,j}}
  \right]\phi_i,
  \\
  \dd\pi_i^a
  &=
  -\frac{1}{2}\dd\log P(a)\,\pi_i^a
  -
  \frac{1}{2}
  \left[
    \sum_{(p,q)\in\{(1,4),(2,3)\}}
    \frac{1}{r_p-r_q}\,
    \dd\log\qty(\frac{a-r_p}{a-r_q})
  \right]\psi_i
  -
  \frac{1}{2}
  \left[
    \sum_{j=1}^{5}
    \frac{r_j\,\dd r_j}{\Delta_{5,j}}
  \right]\phi_i,
\end{align}
where
\begin{equation}
  \Delta_{n,j}
  =
  \prod_{i=1,~i\neq j}^{n}
  (r_j-r_i),
  \qquad
  r_5=a.
\end{equation}
The differential equation for $\pi_i^\infty$ is inhomogeneous with respect
to the $(\psi_i,\phi_i)$ subsystem: it contains no term proportional to
$\pi_i^\infty$. This is consistent with the constant normalization of its
residue in \cref{eq:period-matrix}. By contrast, the residue of
$\omega_{\pi^a}$ is proportional to $P(a)^{-1/2}$, which produces the
homogeneous term in the second equation of \cref{eq:pi-de}. It can be
removed by the algebraic normalization
\begin{equation}
  \widehat{\pi}_i^a
  =
  \sqrt{P(a)}\,\pi_i^a.
\end{equation}

Since the maximal cuts of $\vb{J}_{\mathrm{R},\kappa}$ are represented by
the one-forms $\omega_\kappa$, the period matrix in
\cref{eq:period-matrix} provides a fundamental matrix for the LO
homogeneous diagonal block of their differential equation \cite{Primo:2016ebd,Bosma:2017ens}. After
identifying the elliptic modulus and marked-point parameter, the
equations above reproduce \cref{eq:de-rational-elliptic-0} up to a change of basis that is rational
in the roots. The integrand-level matching in
\cref{eq:eLS-candidates} therefore yields a rational basis
$\vb{J}_{\mathrm{R}}$ whose diagonal differential equation is linear in
$\epsilon$, as in \cref{eq:de-rational-elliptic}.

The universal transformation \cref{eq:TC}, constructed from the elliptic
leading singularities, then brings this differential equation to the
special form of \cref{eq:de-elliptic-C}, in which every matrix element is
proportional either to $\epsilon$ or to $1-n\epsilon$, with integer $n$.
Its LO homogeneous connection is second-order nilpotent, so the
corresponding path-ordered exponential truncates and the solution can be
written order by order in $\epsilon$ in terms of iterated integrals.

\begin{example}
\label{ex:numerator-ansatz}

We now illustrate the construction of the basis $\vb*{J}_{\mathrm{R}}$ for the two-loop sunrise family \texttt{SR} with two distinct masses whose DE we have already discussed in \cref{ex:functions}.
We will use the Baikov parametrization with variables
\begin{equation}
\vb*{z} = \left\{ k_1^2-m^2,\ k_2^2-M^2,\ (k_1-k_2-p)^2-m^2,\ k_1\cdot p,\ k_2\cdot p \right\}  \,,
\end{equation}
where the first three are denominators defining the top sector of the family and the last two are irreducible scalar products. 
The Baikov parametrization in $d=2 - 2 \epsilon$ dimensions reads on the maximal cut
\begin{equation}
  G^{1,1,1,-q_4,-q_5}_{\texttt{SR}} \xrightarrow{\mathcal{P}} C(\epsilon)\; (p^2)^\epsilon \int\limits_{\gammacircle{z_1}\otimes\gammacircle{z_2}\otimes\gammacircle{z_3}} \dd^5\vb*{z} \quad  \mathcal{B}^{-\epsilon}\, \frac{1}{\mathcal{B}} \frac{z_4^{q_4} z_5^{q_5}}{z_1 z_2 z_3}  
  = \int \dd z_4 \wedge \dd z_5 \;  \frac{z_4^{q_4} z_5^{q_5}}{\mathcal{B}_\text{cut}(z_4,z_5)}   \;+\order{\epsilon} \,,
\end{equation}
where $C(\epsilon)$ is an  $\epsilon$-dependent constant that is not relevant here, we set $p^2 = 1$, and
\begin{equation}
 \begin{aligned}
\mathcal{B}_\text{cut}(z_4,z_5)  &= -2 z_4^2 z_5
+2 z_4 z_5^2
-\left(1+M^2\right)z_4^2
+\left(3+M^2\right)z_4z_5
-\left(1+m^2\right)z_5^2
\\
& \qquad
+\left(1+M^2\right)z_4
-\left(1+M^2\right)z_5
+m^2M^2
-\frac{1}{4}\left(1+M^2\right)^2 .
\end{aligned}
\end{equation}
We start by considering an ansatz with powers of numerators chosen such that they cover the integrands in \cref{eq:integrand-basis},
\begin{equation}
  \mathcal{I}[n] = n_1\, G_{\texttt{SR}}^{1,1,1,0,0} + n_2\, G_{\texttt{SR}}^{1,1,1,-1,0} + n_3\, G_{\texttt{SR}}^{1,1,1,0,-1} + n_4\, G_{\texttt{SR}}^{1,1,1,-2,0} + n_5\, G_{\texttt{SR}}^{1,1,1,-1,-1} + n_6\, G_{\texttt{SR}}^{1,1,1,0,-2}  \,,
\end{equation}
and derive its parametrization
\begin{equation}
\mathcal{I}[n] \xrightarrow{\mathcal{P}} \frac{\dd z_4 \wedge \dd z_5}{\mathcal{B}_\text{cut}} \qty(n_1  + n_2\, z_4 + n_3\, z_5 + n_4\, z_4^2 + n_5\, z_4 z_5 + n_6\, z_5^2)  \,.
\end{equation}
We first perform the logarithmic decomposition with respect to $z_4$.  Eliminating non-elliptic double poles requires $n_4=n_5=0$, after which
the ansatz becomes
\begin{equation}
  \begin{aligned}
    \mathcal{I}[n]  &= \frac{1}{2}\frac{n_2}{1+M^2 + 2 z_5} \; \qty[ \dd z_5 \wedge \dd\log\big(\alpha_1(z_4,z_5)\big) ] \\
                    & +\frac{n_1 + \frac{1}{2} n_2 + \frac{1}{2} n_2\,z_5 + n_3\, z_5 + n_6 z_5^2}{\sqrt{P(z_5})} \; \qty[\dd z_5 \wedge \dd\log\big(\beta_1(z_4,z_5)\big)]  \,,
  \end{aligned}
\end{equation}
where the explicit forms of the algebraic functions $\alpha_1$ and $\beta_1$ will not be needed. The elliptic curve is defined by
\begin{equation}
  P(z_5) = \left(z_5+\frac{1+M^2}{2}\right) \left(z_5+\frac{1-4m^2+M^2}{2}\right) \left(z_5+M\right) \left( z_5 - M \right)  \,.
\end{equation}
For
\begin{equation}
  n_1=-1,
  \qquad
  n_2=2,
  \qquad
  n_3=-1,
  \qquad
  n_6=0,
\end{equation}
the elliptic term vanishes and the remaining integrand is a logarithmic
two-form. The corresponding integral combination, however, reduces to
zero by IBP identities. We may therefore set $n_2=0$, leaving
\begin{equation}
    \mathcal{I}[n]  = \frac{n_1 + n_3\, z_5 + n_6 z_5^2}{\sqrt{P(z_5})} \; \qty[\dd z_5 \wedge \dd\log\big(\beta_1(z_4,z_5)\big)]  \,.
\end{equation}
Matching this ansatz to each of $N_\kappa$ in \cref{eq:integrand-basis} gives
\begin{equation}
  \begin{aligned}
    \vb*{J}_{\mathrm{R},\psi}: \qquad  & n^\psi_1=1,\quad n^\psi_{i\neq 1}=0   \,,\\
    \vb*{J}_{\mathrm{R},\phi}: \qquad  & n^\phi_1=2 m^2\,M,\quad n^\phi_{3}=(2 m^2 - M^2-1),\quad n^\phi_6 = -2  \,,\\
    \vb*{J}_{\mathrm{R},\pi^\infty}: \qquad  & n^{\pi^{\infty}}_3=1,\quad n^{\pi^{\infty}}_{i\neq 3}=0  \,,  \\  
  \end{aligned}
\end{equation}
which matches the basis $\vb{J}_\mathrm{R}$ given in \cref{eq:rational_basis_sunrise2m} on the maximal cut.
This example was also considered in ref.~\cite{Duhr:2025lbz}, and the first step of the approach therein agrees with $\vb{J}_\mathrm{R}$, except for $\vb*{J}_{\mathrm{R},\phi}$.

To determine the required subsector completion of
$\vb{J}_{\mathrm{R},\phi}$, in this case it is sufficient to release the cut on $z_2$. The resulting
integrand has a non-vanishing iterated residue at $z_5=\infty$ and
$z_4=\infty$. This residue is cancelled by adding $-G_{\texttt{SR}}^{1,0,1,0,-1}$ ($G_{\texttt{SR}}^{1,0,1,0,-1} = -G_{\texttt{SR}}^{1,0,1,0,0}$ under IBP reduction).

\end{example}

Before we conclude this section, let us comment on the related discussion of elliptic LS in Ref.~\cite{Forner:2026vby}.
We emphasize that the matching in \cref{eq:eLS-candidates} is performed entirely within ordinary algebraic cohomology at fixed integer dimension $d_0$.
In particular, the first-, second-, and third-kind representatives are extracted from the same leading order in $\epsilon$ of the complete integrand ansatz. 
This differs from the generalized notion of LS adopted in Ref.~\cite{Forner:2026vby}, where contributions beyond leading order in $\epsilon$ are considered for the purpose of identifying an $\epsilon$-factorized basis.
Beyond leading order in $\epsilon$, the dimensionally regulated integrand is no longer an ordinary algebraic differential form, and the corresponding quantities are coefficients in the $\epsilon$ expansion of twisted periods. 
In Ref.~\cite{Forner:2026vby}, it is argued that these higher-order contributions can be obtained indirectly from the differential equations and are necessary to establish the connection between the $\epsilon$-factorized (canonical) bases and the corresponding integrands.
Taking into account such higher-order LS would necessarily lead to a non-rational basis, and we therefore do not consider them here.

\subsection{The origin of \texorpdfstring{$n$}{n}}
\label{sec:n-meaning}

In Ref.~\cite{Chaubey:2025adn}, we observed that the differential equation for the basis $\vb{J}_{\mathcal C}$ contains an integer $n$: the $\epsilon$ dependence of each matrix element factorizes either as $\epsilon$ or as $1-n\epsilon$.
We now show that, when the maximal cut admits a one-parameter representation, $n$ is determined by the local $\epsilon$-dependent scaling of the integrand at the double pole associated with the second-kind differential.

In the examples considered here, the required one-parameter representation is obtained using the loop-by-loop Baikov representation~\cite{Frellesvig:2017aai}.
The role of this representation is to retain the complete $\epsilon$ dependence and express it through powers of irreducible polynomials.
After reducing the maximal cut to one remaining integration variable $z$, it takes the form
\begin{equation}
\label{eq:loopbyloop}
  C(\vb*{s},\epsilon)
  \int
  \dd z\,
  \prod_i
  p_i(z,\vb*{s})^{\alpha_i},
\end{equation}
where all factors independent of $z$ have been absorbed into $C(\vb*{s},\epsilon)$ and
\begin{equation}
  \alpha_i
  =
  \frac{1}{2}(a_i+b_i\epsilon),
  \qquad
  a_i,b_i\in\mathbb Z.
\end{equation}
Let $z_\star$ denote the location of the double pole, and define $\nu_i=\operatorname{ord}_{z_\star}(p_i)$ as the order of $p_i$ at $z_\star$, with positive values corresponding to zeros and negative values to poles.
Let $t$ be a local coordinate that vanishes at $z_\star$.
We then have locally
\begin{equation}
  p_i(z,\vb*{s})
  \sim
  c_i(\vb*{s})\,t^{\nu_i}.
\end{equation}
The $\epsilon$-dependent part of the integrand therefore scales as
\begin{equation}
  \prod_i
  p_i(z,\vb*{s})^{b_i\epsilon/2}
  \sim
  t^{\frac{\epsilon}{2}\sum_i b_i\nu_i}
  =
  t^{n\epsilon},
\end{equation}
with
\begin{equation}
\label{eq:n-formula}
  n
  =
  \sum_i\frac{b_i\nu_i}{2}.
\end{equation}
Since the $\epsilon$-independent second-kind differential has a double pole at $t=0$, its regulated local behavior is proportional to $t^{-2+n\epsilon}\dd t$.
The identity
\begin{equation}
  \dd\!\left(t^{-1+n\epsilon}\right)
  =
  -(1-n\epsilon)\,
  t^{-2+n\epsilon}\dd t
\end{equation}
then explains the appearance of the factor $1-n\epsilon$.

Throughout this work, we choose the double pole to lie at $z_\star=\infty$.
If $p_i$ is a polynomial of degree $d_i$, then $\nu_i=-d_i$, and \cref{eq:n-formula} becomes
\begin{equation}
  n
  =
  -\sum_i\frac{b_i d_i}{2}.
\end{equation}
As illustrated below, this local scaling exponent reproduces the integer appearing in the factors $1-n\epsilon$ in \cref{eq:de-elliptic-C}.\footnote{
  We thank Stefan Weinzierl for an early comment that motivated us to investigate this connection.
}

Although we have not considered multivariate representations explicitly, we expect that the relevant double pole can be isolated by taking residues associated, at leading order in $\epsilon$, with a $(k-1)$-form of $\dd\log$ type, as discussed above.
The remaining one-form could then be analyzed locally in the same way as the one-parameter representations considered here.

\begin{example}
\label{ex:d-family-n}

Consider the \texttt{D} integral family in
\cref{tab:examples-summary}. Its loop-by-loop Baikov representation
contains the factor
\begin{equation}
\label{eq:dijet-compact}
  \frac{
    s^{-2\epsilon-1}
    (s+t+z)^{-2\epsilon-1}
  }{
    \sqrt{z(s+z)}
  }
  \left(z^2+sz-4s\right)^{-\epsilon-\frac{1}{2}}.
\end{equation}
The factor $s^{-2\epsilon-1}$ is independent of $z$ and is therefore
absorbed into $C(\vb*{s},\epsilon)$. The remaining irreducible polynomials
are
\begin{equation}
  p_1(z)=z,
  \qquad
  p_2(z)=s+z,
  \qquad
  p_3(z)=z^2+sz-4s,
  \qquad
  p_4(z)=s+t+z,
\end{equation}
with degrees
\begin{equation}
  d_1=d_2=d_4=1,
  \qquad
  d_3=2.
\end{equation}
Their exponents are
\begin{equation}
  \alpha_1=\alpha_2=-\frac{1}{2},
  \qquad
  \alpha_3=-\epsilon-\frac{1}{2},
  \qquad
  \alpha_4=-2\epsilon-1.
\end{equation}
Writing $\alpha_i=\tfrac{1}{2}(a_i+b_i\epsilon)$ gives
\begin{equation}
  b_1=b_2=0,
  \qquad
  b_3=-2,
  \qquad
  b_4=-4.
\end{equation}
Substitution into \cref{eq:n-formula} then yields
\begin{equation}
  n
  =
  -\frac{1}{2}
  \left[
    (0)(1)+(0)(1)+(-2)(2)+(-4)(1)
  \right]
  =
  4,
\end{equation}
in agreement with the factor $1-4\epsilon$ appearing in the differential
equation for this family.

\end{example}

We have verified that \cref{eq:n-formula} reproduces the value of $n$ for
all examples considered in Ref.~\cite{Chaubey:2025adn}.

We conclude this section by relating the interpretation of $n$ to two
alternative constructions of the second-kind candidate. First, it clarifies
the connection with the approach of Ref.~\cite{Chen:2025hzq}. In our
construction, the double pole is placed at infinity in the original Baikov
variable. Consequently, all polynomial factors in
\cref{eq:loopbyloop} contribute to $n$, leading to the sum in
\cref{eq:n-formula}. In Ref.~\cite{Chen:2025hzq}, the map to Legendre form
instead sends a selected singular divisor associated with one of the
factors $p_i(z)$ to infinity. The sum then reduces to the contribution from
that factor,
\begin{equation}
  n=-\frac{1}{2}b_i d_i,
\end{equation}
where $d_i$ denotes its scaling degree under the chosen map. In particular,
$n$ may vanish. In \cref{ex:d-family-n}, choosing the
double pole to correspond to $z=-s$, rather than to infinity, selects the
factor $p_2(z)=z+s$, for which $b_2=0$. We then obtain a differential
equation of the form \cref{eq:de-elliptic-C} with $n=0$, in agreement with
Ref.~\cite{Chen:2025hzq}.

The same interpretation explains why using
$\partial_{s_a}\vb{J}_{\mathrm{R},\psi}$ as the second-kind candidate,
instead of constructing $\vb{J}_{\mathrm{R},\phi}$ by the method described
above, can lead to more complicated $\epsilon$ dependence in the
homogeneous differential equation. Depending on the chosen kinematic
variable $s_a$, the derivative may produce a
$\mathbb{Q}(\vb*{s},\epsilon)$-linear combination of integrands with double
poles at several locations. These poles may carry different
$\epsilon$-dependent scaling exponents, which cannot in general be
normalized simultaneously. By contrast, the construction above selects a
representative with a double pole at a single chosen location and therefore
gives rise to a single factor $1-n\epsilon$.

\section{Off-diagonal differential equation blocks}
\label{sec:inh-de}

Off-diagonal DE blocks involving an elliptic sector arise in two distinct situations.
They may describe the dependence of the elliptic sector on its proper subsectors, or the dependence of higher sectors on an elliptic sector that appears among their subsectors.
In this section, we consider only the former case and leave the latter for future work.

As discussed in \cref{sec:elliptic-case}, a common approach to elliptic
Feynman integrals is to construct a basis satisfying an
$\epsilon$-factorized differential equation with constant intersection
matrix. Typically, one first transforms the diagonal elliptic block using
a monodromy-dependent transformation
$T_{\mathrm{hom}}(\vb*{s},\epsilon,\vb*{E})$. The off-diagonal blocks are
then treated by triangular transformations involving master integrals from
proper subsectors,
\begin{equation}
\label{eq:inh-de-ep-fact}
  \IefactE
  =
  \left(
    \prod_{\sigma\in\text{subsectors}}
    T_\sigma(\vb*{s},\epsilon,\vb*{E}^{\prime})
  \right)
  T_{\mathrm{hom}}(\vb*{s},\epsilon,\vb*{E})\,
  \vb{I}_{\mathrm{R}},
\end{equation}
where $\vb*{E}^{\prime}\supseteq\vb*{E}$ may contain additional functions
with non-trivial monodromy. If the objective is only to obtain an
$\epsilon$-factorized differential equation, the functions required to
solve the inhomogeneous equations can be included in $\vb*{E}^{\prime}$.
This procedure does not, however, identify a preferred rational basis
before these functions are introduced.

In our approach the issue arises because the maximal-cut construction of
\cref{sec:recap-method} determines an elliptic top-sector integral only
modulo integrals from proper subsectors, whose maximal cuts vanish. The
off-diagonal blocks therefore contain the information needed to choose
representatives of the maximal-cut candidates away from the cut. To obtain
a function basis that separates the rational and multivalued parts of the
amplitude, this ambiguity must be resolved through rational subsector
shifts before introducing monodromy-dependent transformations.

To make this ambiguity explicit, we separate the elliptic sector from its
proper subsectors and write the rational differential equation
schematically as
\begin{equation}
\label{eq:rational-de-blocks}
  \dd
  \begin{pmatrix}
    \vb{J}_{\mathrm{R},\mathrm{ell}}\\
    \vb{J}_{\mathrm{R},\mathrm{sub}}
  \end{pmatrix}
  =
  \begin{pmatrix}
    M_{\mathrm{ell}} & D\\
    0 & M_{\mathrm{sub}}
  \end{pmatrix}
  \begin{pmatrix}
    \vb{J}_{\mathrm{R},\mathrm{ell}}\\
    \vb{J}_{\mathrm{R},\mathrm{sub}}
  \end{pmatrix}.
\end{equation}
A rational change of basis that leaves the maximal cuts
unchanged has the form
\begin{equation}
\label{eq:rational-subsector-shift-matrix}
  \vb{J}_{\mathrm{R},\mathrm{ell}}^{\prime}
  =
  \vb{J}_{\mathrm{R},\mathrm{ell}}
  +
  R(\vb*{s},\epsilon)\,
  \vb{J}_{\mathrm{R},\mathrm{sub}},
  \qquad
  R_{i j}(\vb*{s},\epsilon)
  \in
  \mathbb{Q}(\vb*{s},\epsilon).
\end{equation}
Under this transformation, the off-diagonal block becomes
\begin{equation}
\label{eq:off-diagonal-transformation}
  D^{\prime}
  =
  D+\dd R+R M_{\mathrm{sub}}-M_{\mathrm{ell}}R.
\end{equation}
The problem addressed in this section is to choose a rational $R$ for
which $D^{\prime}$ has a controlled dependence on $\epsilon$.

It is instructive to recall how the analogous ambiguity is resolved at the integrand level in the algebraic case. 
Integrals whose maximal cuts have $\dd\log$ integrands may
develop additional double poles or non-integer LS when cut conditions are released. These
are removed by subtracting rational linear combinations of subsector
master integrals, after which the off-diagonal blocks are
$\epsilon$-factorized. The algebraic transformation that removes the LO
homogeneous block in
\cref{eq:de-rational-sqrt-hom,eq:de-rational-nested-sqrt} can then be
applied after the rational subsector dependence has been fixed.
The corresponding question in the elliptic case is how to complete the
maximal-cut candidates
$\vb{J}_{\mathrm{R},\psi}$,
$\vb{J}_{\mathrm{R},\phi}$, and
$\vb{J}_{\mathrm{R},\pi^a}$
by rational subsector contributions.

In Ref.~\cite{Chaubey:2025adn}, we proposed that the construction reviewed
in \cref{sec:recap-method} could be extended beyond the maximal cut by
successively releasing cut conditions. By analogy with the $\dd\log$
construction in the algebraic case, it was expected that suitable rational
subsector subtractions would lead to $\epsilon$-factorized off-diagonal
blocks. Parametrizations other than Baikov may be more convenient for this
purpose. In particular, using the momentum-space parametrization of
Ref.~\cite{Henn:2020lye} without imposing any cuts, we reproduced for the
\texttt{SR} family the rational basis given in
\cref{eq:rational_basis_sunrise2m}.
Our more general examples show that this picture does not extend
unchanged. Consistently with the observations of
Refs.~\cite{Yang:2025ofz,Forner:2026vby}, the integrand construction for
$\vb{J}_{\mathrm{R},\psi}$ and $\vb{J}_{\mathrm{R},\pi^a}$ can be carried
out after releasing the maximal-cut conditions. Unwanted contributions can
then be subtracted at LO in $\epsilon$, much as in the $\dd\log$ case. In
the examples considered here, such corrections are required only for a
small number of subsectors.

The situation is more complicated for
$\vb{J}_{\mathrm{R},\phi}$. Its integrand must retain the second-kind double
pole in the final integration variable associated with the elliptic curve,
while avoiding unwanted higher-order poles in the remaining variables.
We can implement this requirement completely for families with a simple
subsector hierarchy, such as the \texttt{SR} family, but we do not find a
general integrand-level prescription for more complicated examples.
Moreover, even when suitable subtractions can be identified, the resulting
off-diagonal blocks need not have the expected $\epsilon$ dependence.
Leaving $\vb{J}_{\mathrm{R},\phi}$ uncorrected typically produces an even
more complicated dependence on $\epsilon$.
We therefore adopt a different strategy for
$\vb{J}_{\mathrm{R},\phi}$ and address the problem at the level of
the differential equations: we determine which simplifications of the
off-diagonal blocks can be achieved through rational subsector
transformations.

In the following, we assume that the unwanted contributions in
$\vb{J}_{\mathrm{R},\psi}$ and $\vb{J}_{\mathrm{R},\pi^a}$ have been
removed at the integrand level, as in the logarithmic case.
Then, for each proper subsector $\sigma$, we consider a general shift of
$\vb{J}_{\mathrm{R},\phi}$ by rational linear combinations of its master
integrals,
\begin{equation}
\label{eq:rational-subsector-ansatz}
  \vb{J}_{\mathrm{R},\phi}^{\prime}
  =
  \vb{J}_{\mathrm{R},\phi}
  +
  \sum_{i\in\sigma}
  r_i(\vb*{s},\epsilon)\,
  \vb{J}_{\mathrm{R},i},
  \qquad
  r_i(\vb*{s},\epsilon)
  \in
  \mathbb{Q}(\vb*{s},\epsilon),
\end{equation}
and study the resulting off-diagonal DE blocks. 
See e.g.~refs.~\cite{Gehrmann:2014bfa,Meyer:2017joq} for a similar approach in the context of rational canonical differential equations.

For the family \texttt{SR}, the other sunrise families studied in
Ref.~\cite{Chaubey:2025adn}, and the three-point elliptic sector of the
family \texttt{D}, rational shifts of the form
\cref{eq:rational-subsector-ansatz} can readily produce
$\epsilon$-factorized off-diagonal blocks. For the full family \texttt{D}
and the family \texttt{TT}, however, we find that no rational shift of this
form yields complete $\epsilon$ factorization. The existence of such a
shift can be decided by substituting
\cref{eq:rational-subsector-ansatz} into the differential equation. This
produces a system of linear partial differential equations for
$r_i(\vb*{s},\epsilon)$, for which one can either construct a rational
solution or establish that none exists
\cite{BARKATOU1999547,ABRAMOV19897}. In practice, we use the corresponding
implementation in \texttt{Maple}~\cite{MaplesoftRationalSolution}.

The central observation is that the obstruction to complete
$\epsilon$ factorization does not lead to arbitrary $\epsilon$ dependence.
In every family and subsector considered in this work, we can always find a
rational shift for which the off-diagonal entries take the same restricted
form
\begin{equation}
\label{eq:rational-subsector}
\begin{aligned}
  (\dd\vb{J}_{\mathrm{R},\psi})_i
  &=
  \epsilon\,\omega_{\psi,i},
  \\
  (\dd\vb{J}_{\mathrm{R},\phi}^{\prime})_i
  &=
  \frac{\epsilon}{1-n\epsilon}\,\omega_{\phi,i}^{1}
  +
  \frac{\epsilon^2}{1-n\epsilon}\,\omega_{\phi,i}^{2},
  \\
  (\dd\vb{J}_{\mathrm{R},\pi^a})_i
  &=
  \epsilon\,\omega_{\pi^a,i},
\end{aligned}
\end{equation}
where $(\dd\vb{J}_{\mathrm{R},\kappa})_i$ denotes the coefficient of the
subsector master integral $\vb{J}_{\mathrm{R},i}$, and the $\omega$ are
rational one-forms. All other off-diagonal entries can be made
$\epsilon$-factorized.
We further find that the remaining rational freedom can be used to impose
$\big(\partial_{s_a}\vb{J}_{\mathrm{R},\psi}\big)_i=0$ for a chosen
kinematic variable $s_a$. When the elliptic curve depends on only one
kinematic variable, this condition sets the entire one-form
$\omega_{\psi,i}$ to zero, and \cref{eq:rational-subsector} reduces to
\begin{equation}
\label{eq:rational-subsector-2}
\begin{aligned}
  (\dd\vb{J}_{\mathrm{R},\psi})_i
  &=
  0,
  \\
  (\dd\vb{J}_{\mathrm{R},\phi}^{\prime})_i
  &=
  \frac{\epsilon^2}{1-n\epsilon}\,\omega_{\phi,i},
  \\
  (\dd\vb{J}_{\mathrm{R},\pi^a})_i
  &=
  \epsilon\,\omega_{\pi^a,i}.
\end{aligned}
\end{equation}

The form in \cref{eq:rational-subsector} is obtained separately for each
subsector as follows.
\begin{enumerate}
  \item
  We introduce the rational ansatz
  \cref{eq:rational-subsector-ansatz}, substitute it into the differential
  equation, and compute the transformed off-diagonal block using
  \cref{eq:off-diagonal-transformation}.

  \item
  We expand the rational coefficients $r_i(\vb*{s},\epsilon)$ around
  $\epsilon=0$ and require all coefficients of order $\epsilon^k$, with
  $k\geq2$, in
  $(\dd\vb{J}_{\mathrm{R},\psi})_i$ and
  $(\dd\vb{J}_{\mathrm{R},\pi^a})_i$ to vanish. These conditions generate
  recurrence relations for the coefficients in the $\epsilon$ expansion
  of $r_i(\vb*{s},\epsilon)$. The recurrences can be resummed into rational
  functions of $\epsilon$, fixing the complete $\epsilon$ dependence of
  $r_i$ in terms of at most two undetermined rational functions of
  $\vb*{s}$.

  \item
  We require the $\epsilon^0$ terms in
  $(\dd\vb{J}_{\mathrm{R},\psi})_i$ and
  $(\dd\vb{J}_{\mathrm{R},\pi^a})_i$ to vanish. The corresponding entries
  then become proportional to $\epsilon$, while the
  $\vb{J}_{\mathrm{R},\phi}^{\prime}$ entry acquires the restricted
  dependence shown in \cref{eq:rational-subsector}.
\end{enumerate}
These conditions determine the remaining functions up to a single rational
function. When such a function remains, we fix it by imposing
$\big(\partial_{s_a}\vb{J}_{\mathrm{R},\psi}\big)_i=0$. As discussed
above, when the elliptic curve depends on a single kinematic variable, this
gives the form in \cref{eq:rational-subsector-2}. In multivariate cases,
the same condition may also set the remaining partial derivatives of
$\vb{J}_{\mathrm{R},\psi}$ to zero, but need not do so. 
Finding a better motivated criterion for fixing this freedom is left for future work.

Interestingly, the structure in \cref{eq:rational-subsector} appears to be
insensitive to whether the set $\vb*{E}$ must be enlarged beyond the
maximal cut. For example, in the \texttt{TT} family, an additional
monodromy-dependent function is required to $\epsilon$-factorize certain
off-diagonal DE blocks \cite{Becchetti:2025oyb}.
Nevertheless, the rational basis continues to satisfy
\cref{eq:rational-subsector}. The enlargement of $\vb*{E}$ therefore
affects the subsequent non-rational transformation to an
$\epsilon$-factorized basis, but not the restricted $\epsilon$ dependence
of the rational differential equation.

We emphasize that the construction above is performed entirely within the
rational basis. The $\epsilon$-independent non-rational transformations
that remove the LO homogeneous block, such as those in
\cref{eq:algebraic-basis-sqrt-simple,eq:Tsqrt-nested,eq:TC}, do not alter
the restricted $\epsilon$ dependence in
\cref{eq:rational-subsector}. For example, applying
$T_{\mathcal C}$ gives
\begin{equation}
\label{eq:TC-one-forms}
  (\dd\vb{J}_{\mathcal{C},\phi}^{\prime})_i
  =
  \left(
    \psi\,\omega_{\phi,i}^{1}
    -
    \phi\,\omega_{\psi,i}
  \right)
  \frac{\epsilon}{1-n\epsilon}
  +
  \left(
    \psi\,\omega_{\phi,i}^{2}
    +
    n\phi\,\omega_{\psi,i}
  \right)
  \frac{\epsilon^2}{1-n\epsilon}.
\end{equation}
The basis $\vb{J}_{\mathcal C}$ therefore admits an iterated-integral
solution whose kernels are obtained directly from the rational one-forms
in \cref{eq:rational-subsector} through the transformation
$T_{\mathcal C}$.

The restricted $\epsilon$ dependence is also consistent with the transformations to $\epsilon$-factorized homogeneous blocks proposed in Refs.~\cite{Chen:2025hzq,Yang:2025ofz}. 
Rescaling $\vb{J}_{\mathrm{R},\phi}^{\prime}$ by $(1-n\epsilon)/\epsilon$ generally introduces a $1/\epsilon$ term into the homogeneous block, which must subsequently be removed by a non-rational transformation. 
At the same time, its off-diagonal entry in \cref{eq:rational-subsector} becomes $\omega_{\phi,i}^{1}+\epsilon\,\omega_{\phi,i}^{2}$ and is therefore linear in $\epsilon$. 
With this observation, we also provide evidence that the Laurent-polynomial dependence on $\epsilon$ proposed in Ref.~\cite{e-collaboration:2025frv} extends beyond the maximal cut.
Interestingly, the rational basis for the elliptic sector of the family \texttt{TT} in Ref.~\cite{Badger:2024fgb} was chosen heuristically to simplify the DE, and its relation to our basis is non-trivial.
Nevertheless, after rescaling $\vb{J}_{\mathrm{R},\phi}$ by $1-2\epsilon$, the $\epsilon$ dependence of both the diagonal elliptic block and the off-diagonal blocks involving subsectors agrees with the structure reported in Ref.~\cite{Badger:2024fgb}.

\section{Conclusions and outlook}

In this work, we discussed in more detail the method for constructing rational bases for elliptic Feynman integrals proposed in Ref.~\cite{Chaubey:2025adn}.
The method systematically derives rational bases for multiscale elliptic Feynman integrals that incorporate information about their elliptic geometry.
We argued that such bases provide a natural starting point for constructing transcendental function bases that cleanly separate the rational and multivalued parts of scattering amplitudes, a property that is generally obscured in an $\epsilon$-factorized basis beyond genus-zero geometry.

Extending the analysis of Ref.~\cite{Chaubey:2025adn}, we explained how to predict the integer $n$ appearing in the resulting DEs and proposed a practical approach for fixing the subsector dependence of the rational basis.
We considered the off-diagonal DEs in several examples and found that some of them are too simple to capture the general case.
In particular, examples in which the elliptic curve depends on a single kinematic variable do not capture all features of the multivariate case.
This emphasizes the importance of studying genuinely multiscale examples.

In summary, we envisage the following approach to multiscale amplitudes: use the rational basis proposed in this work to define the subspace of monodromy-invariant transcendental functions relevant for representing the amplitude, and use an $\epsilon$-factorized basis with constant intersection matrix to express these functions in terms of independent iterated integrals to construct an explicit basis for this subspace.
We expect this approach to facilitate both the analytic computation of multiscale amplitudes involving elliptic Feynman integrals through reconstruction techniques and the numerical solution of the corresponding rational DEs.

Much remains to be understood.
Most urgently, it is important to clarify to what extent the residual freedom in the rational off-diagonal blocks is significant and whether there is a more intrinsic principle for fixing it.
It would also be interesting to understand how the method can be extended to cases in which the integrand form considered here does not exist.
Finally, generalizing the construction to higher-genus geometries remains an important direction for future work.

\begin{acknowledgments}

  We thank Cathrin Semper and Sven Stawinski for discussions regarding independence of iterated integrals, as well as Stefan Weinzierl, Yiyang Zhang, and Pau Petit Rosas for discussions on the origin of $n$. 
  We thank Matteo Becchetti, Dmitry Chicherin, Christoph Dlapa, and Simone Zoia for inspiring discussions at the initial stages of this work.
  We are also grateful to Gideon Baur for clarifications regarding the package \texttt{IterInt}.
  V.S.\ extends gratitude to the Bethe Center for Theoretical Physics at the University of Bonn for hospitality during the preparation of this manuscript. E.C.
would like to thank the Physics Institute, University of
Zurich for hospitality.
  We thank Simone Zoia, Lorenzo Tancredi, Sebastian Pögel, Claude Duhr, and Federico Gasparotto for valuable feedback on the manuscript.

  V.S.\ work is supported by the European Research Council (ERC) under the European Union's Horizon 2020 research and innovation programme grant agreement 101019620 (ERC Advanced Grant TOPUP), 
  and by the ERC grant 101220457 ``HiNPrecise''.
  The work of E.C.\ is funded by the ERC grant 101043686 ‘LoCoMotive’. 
  Views and opinions expressed are however those of the author(s) only and
  do not necessarily reflect those of the European Union or the European Research Council.
  Neither the European Union nor the granting authority can be held responsible for them.
\end{acknowledgments}

\appendix
\input{include/appendix.tex}

\bibliography{bibliography.bib}

\end{document}

%% file: include/appendix.tex
\section{Solution through iterated integrals for basis $\vb{J}_{\mathcal{C}}$}
\label{sec:iint-sol}

A universal and minimal transformation by leading singularities (eLS) allows one to write the solution of the differential equation directly in terms of iterated integrals. The formal solution of the differential equation is always given by a path-ordered exponential. For this expansion in $\epsilon$ to be useful, it must truncate order-by-order in $\epsilon$, and the necessary condition for this is that the LO DE matrix be nilpotent. As mentioned earlier, the matrix $M^{(0)}_\mathcal{C}$ is indeed nilpotent, so the path-ordered exponential in \cref{eq:path-exp-sol} truncates, and the solutions can be written explicitly in terms of iterated integrals.

We demonstrate this using the example of the three-point elliptic topology, an elliptic subsector of family $\texttt{D}$, which contributes, for instance, to dijet and diphoton production at the LHC at two loops \cite{vonManteuffel:2017hms,Becchetti:2025rrz,Coro:2025vgn,Ahmed:2025osb,Ahmed:2024tsg}. 
We construct the basis for this family using the algorithm of \cite{Chaubey:2025adn}. The solutions are then written explicitly in terms of this basis, which we denote by $\vb{J}_{\mathcal{C},\psi}$ and $\vb{J}_{\mathcal{C},\phi}$. For scattering amplitude applications one needs the $\epsilon$-expansion of these integrals to higher orders; we write this expansion as $\vb{J}_{\mathcal{C},\psi}=\sum_{j=0} \vb{J}_{\mathcal{C},\psi}^{(j)}\,\epsilon^j$ and $\vb{J}_{\mathcal{C},\phi}=\sum_{j=0} \vb{J}_{\mathcal{C},\phi}^{(j)}\,\epsilon^j$, where we assume that the first non-vanishing contribution to each integral starts at $j=0$.

For completeness, we provide the definition of this integral family, together with the basis that we constructed, in the ancillary files. Note that, for this topology, a basis where the subsectors are fully $\epsilon$-factorized is possible using the algorithm of \cite{Chaubey:2025adn}; nevertheless, we prefer the form given below for consistency with the other topologies discussed in this paper, as explained in section~\ref{sec:inh-de}.

The differential equation for the two top-sector integrals takes the form
\begin{equation}
\begin{aligned}
\left(\begin{array}{c}
    \dd \vb{J}_{\mathcal{C},\psi} \\
    \dd \vb{J}_{\mathcal{C},\phi}
\end{array}
\right)=\;&
\left(
\begin{array}{cc}
 -\dfrac{2\,\epsilon\,\phi\, \dd s}{s(s+16)\psi} & \dfrac{(1-4\epsilon)\dd s}{s^{2}(s+16)\psi^{2}} \\[3ex]
 \dfrac{\epsilon\,\phi\big(\phi-(s+8)\psi\big) \dd s}{s+16} & \dfrac{2\epsilon\big(\phi-(s+8)\psi\big) \dd s}{s(s+16)\psi} \\
\end{array}
\right)
\left(\begin{array}{c}
    \vb{J}_{\mathcal{C},\psi} \\
    \vb{J}_{\mathcal{C},\phi}
\end{array}
\right) \\[2ex]
&+
\left(\begin{array}{c}
    0 \\[1ex]
    \dfrac{\epsilon^2\,\dd s\,\psi}{4\epsilon-1}\left(\vb{J}_1+\dfrac{13}{4}\vb{J}_2+\dfrac{7}{4}\vb{J}_{3}-\dfrac{1}{2}\vb{J}_{4}-\dfrac{3\sqrt{s}}{4\sqrt{s+4}}\vb{J}_{5}\right)
\end{array}
\right)
\end{aligned}
\end{equation}
where the integrals $\vb{J}_{1}$, $\vb{J}_{2}$, $\vb{J}_{3}$, $\vb{J}_{4}$, and $\vb{J}_{5}$ are subsector contributions whose explicit definitions are provided in the ancillary files. Like the top-sector integrals, these also admit an $\epsilon$-expansion,
\[
\vb{J}_k=\sum_{j=0}^{\infty}\vb{J}_k^{(j)}\,\epsilon^j,
\qquad
k\in\{1,2,3,4,5\}.
\]

For illustrative purposes, we present below only those contributions that are required to obtain, for instance, the finite remainder of diphoton production at two loops. We solve the system order by order in $\epsilon$. After imposing the boundary conditions at $s=0$, the only non-vanishing contributions arise at order $\epsilon^4$. More precisely, the only non-zero coefficients are $\vb{J}_{\mathcal{C},\psi}^{(4)}$ and $\vb{J}_{\mathcal{C},\phi}^{(4)}$, which are given by
\[
\begin{aligned}
  \vb{J}_{\mathcal{C},\psi}^{(4)}
&= \int \omega_{\psi_1} \, \vb{J}_{\mathcal{C},\phi}^{(4)}
= \frac{5}{4}\,I_{\gamma}\big(\omega_1,\omega_1,\omega_{\psi_2},\omega_{\psi_1} \big)+I_{\gamma}\big(\omega_2,\omega_3,\omega_{\psi_3},\omega_{\psi_1}\big), \\[1ex]
\vb{J}_{\mathcal{C},\phi}^{(4)}
&= -\int \omega_{\psi_2}\,J_1^{(2)} -\frac{13}{4}\int \omega_{\psi_2}\,J_2^{(2)} +\int \omega_{\psi_3}\,J_{5}^{(2)}
= \frac{5}{4}\,I_{\gamma}\big(\omega_1,\omega_1,\omega_{\psi_2}\big) +I_{\gamma}\big(\omega_2,\omega_3,\omega_{\psi_3}\big)
\end{aligned}
\]
with
\begin{align}
\omega_1 &= \dd \log\left(\frac{-s-\sqrt{s-4}\sqrt{s}}{\sqrt{s-4}\sqrt{s}-s}\right), &
\omega_2 &= \dd \log(s), &
\omega_3 &= \dd \log\left(\frac{s-\sqrt{s}\sqrt{s+4}}{s+\sqrt{s}\sqrt{s+4}}\right), \nonumber \\
\omega_{\psi_1} &= \frac{\dd s}{s^2(16+s)\psi(s)^2}, &
\omega_{\psi_2} &= \dd s\,\psi, &
\omega_{\psi_3} &= \dd s\,\frac{3\sqrt{s}\, \psi}{4\sqrt{s+4}}.
\end{align}

In the second equality, we have substituted the explicit expressions for $\vb{J}_1$, $\vb{J}_2$, and $\vb{J}_{5}$ to obtain the corresponding iterated-integral representation. As is evident from the expressions above, all coefficients below order $\epsilon^4$ vanish for this pair of master integrals.

\section{Representation of elliptic leading singularities through complete elliptic integrals.}
\label{sec:complete-eli}

The elliptic leading singularities $\psi$,  $\phi$, $\pi ^{a_i}$ can be represented in closed form through complete elliptic integrals of first, second, and third kind respectively.
An explicit form depends on the arrangement of roots (see e.g.~\cite{Broedel:2019hyg}). 
For reference we give here a representation for four real roots ordered as $r_1 < r_2 <r_3<r_4$.

Here we employ the following conventions for the complete elliptic integrals,
\begin{align}
  \mathrm{K}(m) &= \int_0^{\frac{\pi}{2}} \frac{\dd\theta}{\sqrt{1 - m \sin^2\theta}}, \\
  \mathrm{E}(m) &= \int_0^{\frac{\pi}{2}} \sqrt{1 - m \sin^2\theta} \, \dd\theta, \\
  \Pi(n, m) &= \int_0^{\frac{\pi}{2}} \frac{\dd\theta}{(1 - n \sin^2\theta)\sqrt{1 - m \sin^2\theta}},
\end{align}
and denote
\begin{equation}
  k =  \frac{r_{32} \, r_{41}}{r_{31} \, r_{42}}, \qquad \bar{k} = 1 - k =  \frac{r_{21} \, r_{43}}{r_{31} \, r_{42}}.
\end{equation}

The integration over the first cycle $\gamma_1$ yields
\begin{align}
  \psi_1 &= \frac{2 \, \mathrm{K}\left( k \right)}{\sqrt{r_{31} \, r_{42}}},\\
  \phi_1 &= 2 \, \sqrt{r_{31} \, r_{42}} \, \mathrm{E}\left( k \right) ,\\
  \pi_1^{\infty} &=  \frac{2}{\sqrt{r_{31} \, r_{42}}} \qty[ r_1 \, \mathrm{K}(k)+ r_{21}\,\Pi\left( \frac{r_{23}}{r_{13}}, k \right)] \, , \\
  \pi_1^{a} &= \frac{2}{\sqrt{r_{31} \, r_{42}}}\frac{1}{r_1 - a} \qty[\mathrm{K}\left( k \right)-  \frac{r_{21}}{(r_2 - a) }\,\Pi\left( \frac{(r_1-a) \, r_{23}}{(r_2 - a) \, r_{13}}, k \right)]  \,.
\end{align}
And the integration over the second cycle $\gamma_2$ yields
\begin{align}
  \psi_2 &= \frac{-2\,\mathrm{i} \, \mathrm{K}(\bar{k})}{\sqrt{r_{31} \, r_{42}}}  \,, \\
  \phi_2 &= -2\,\mathrm{i}\sqrt{r_{31}r_{42}}\, \left[\mathrm{K}(\bar{k})-\mathrm{E}(\bar{k})\right] \,, \\
  \pi_2^{\infty} &= \frac{-2\,\mathrm{i}}{\sqrt{r_{31}r_{42}}} \left[ r_2\,\mathrm{K}(\bar{k}) + r_{32}\,\Pi\!\left(\frac{r_{34}}{r_{24}},\bar{k}\right) \right] \,, \\
  \pi_2^{a} &= \frac{-2\,\mathrm{i}}{(r_2-a)\sqrt{r_{31}r_{42}}}
  \left[ \mathrm{K}(\bar{k}) - \frac{r_{32}}{r_3-a}\, \Pi\!\left( \frac{(r_2-a)\,r_{34}}{(r_3-a)\,r_{24}}, \bar{k} \right) \right]\,.
  \end{align}


\section{Definitions of one-forms in \cref{ex:functions}}
\label{sec:rat-one-form-aux}

Using the following denominators,
\begin{equation}
  \begin{aligned}
    q_{1}&=\frac{1}{m}, &
    q_{2}&=\frac{1}{2 m-M-1}, &
    q_{3}&=\frac{1}{2 m-M+1}, &
    q_{4}&=\frac{1}{M-1}, \\
    q_{5}&=\frac{1}{M}, &
    q_{6}&=\frac{1}{M+1}, &
    q_{7}&=\frac{1}{2 m+M-1}, &
    q_{8}&=\frac{1}{2 m+M+1},
  \end{aligned}
\end{equation}
it is convenient to introduce the recurring denominator products
\begin{equation}
  \begin{aligned}
    Q_A &= q_{1}q_{2}q_{3}q_{4}q_{6}q_{7}q_{8}, &
    Q_B &= q_{5}\,Q_A, &
    Q_C &= q_{3}q_{6}q_{7}, \\
  \end{aligned}
\end{equation}
The rational one-forms are then
\small
\begin{equation}
  \begin{aligned}
    \omega_{1}  &=  -2 \dd{m} q_{1}-2 \dd{M} q_{5} \,,\\
    \omega_{2}  &=  -4 \dd{m} q_{1} \,,\\
    \omega_{3}  &=  Q_A \left(4 \left(M^2-1\right) \dd{m} \left(4 m^2-M^2-1\right)+4 m M \dd{M} \left(-4 m^2+M^2+3\right)\right) \,,\\
    \omega_{4}  &=  Q_C \left(-4 m (M+1) \dd{m}-2 \dd{M} \left(2 m^2-M^2+M\right)\right) \,,\\
    \omega_{5}  &=  Q_B \left(m \dd{M} \left(-8 m^2+6 M^2+2\right)-4 M \left(M^2-1\right) \dd{m}\right) \,,\\
    \omega_{6}  &=  Q_B \left(4 M \left(M^2-1\right) \dd{m} \left(-10 m^2+M^2+1\right)+4 m \left(m^2-1\right) \dd{M} \left(4 m^2+5 M^2-1\right)\right) \,,\\
    \omega_{7}  &=  Q_C \left(8 m M (M+1) \dd{m}-4 M \dd{M} \left(2 m^2+M-1\right)\right)  \,,\\
    \omega_{8}  &=  Q_C \left(-4 m (M-1) M \left(M^2-1\right) \dd{m}-2 m^2 (M-1) \dd{M} \left(4 m^2-3 M^2-1\right)\right) \,,\\
    \omega_{9}  &=  Q_C \left(-24 m M (M+1) \dd{m} \left(m^2-M\right)-6 \dd{M} \left(4 m^4 (M-2)+m^2 \left(2-2 (M-2) M^2\right)+(M-1) M \left(M^2+1\right)\right)\right)  \,,\\
    \omega_{10} &=  Q_C \left(4 m (M+1) \dd{m} \left(2 m^2+(M-4) M+1\right)-4 \left(m^2-1\right) \dd{M} \left(2 m^2+(M-1) M\right)\right)  \,,\\
    \omega_{11} &=  Q_A \left(8 \left(M^2-1\right) M^2 \dd{m}+4 m M \dd{M} \left(4 m^2-3 M^2-1\right)\right) \,,\\
    \omega_{12} &=  Q_A \left(12 M \left(M^2-1\right) \dd{m} \left(12 m^4-3 m^2 \left(M^2+1\right)+M^3+M\right) \right. \\
                &  \qquad \left.- 6 m M \dd{M} \left(24 m^4 M-2 m^2 (M (M (3 M+2)+9)+2)+3 M^4+4 M^2+1\right)\right)  \,,\\
    \omega_{13} &=  Q_A \left(-2 \left(M^2-1\right) \dd{m} \left(4 m^2-M^2-1\right) \left(10 m^2-M^2-1\right)-4 m M \dd{M} (m-M) (m+M) \left(4 m^2-M^2+5\right)\right) \,,\\
  \end{aligned}
\end{equation}
\normalsize
and
\small
\begin{equation}
  \begin{aligned}
    \bar{\omega}_1 &= q_{1}q_{5}(-2 M \dd{m}-2 m \dd{M}), \\
    \bar{\omega}_2 &= -4 \dd{m}\, q_{1}, \\
    \bar{\omega}_3 &= Q_A \Big(4 m M \dd{M} \big(16 m^4 (M^2-3)-8 m^2 (M^4-3 M^2-5)+M^6-15 M^4-3 M^2-7\big) \\
    &\qquad -4 (M^2-1) \dd{m} \big(16 m^4 (M^2+1)-8 m^2 (M^4+3 M^2+1)+M^6-7 M^4-7 M^2+1\big)\Big), \\
    \bar{\omega}_4 &= Q_B \Big(16 M \dd{m} \big(-4 m^2 (M^4-1)+M^6+6 M^4-6 M^2-1\big) \\
    &\qquad -4 m \dd{M} \big(16 m^4 (M^2+1)-8 m^2 (3 M^4+9 M^2+1)+5 M^6+45 M^4+21 M^2+1\big)\Big), \\
    \bar{\omega}_5 &= Q_B \Big(m \dd{M} \big(-64 m^6 (M^2+1)^2+16 m^4 (5 M^6+17 M^4+15 M^2+3) \\
    &\qquad\qquad -4 m^2 (7 M^8+28 M^6+58 M^4+36 M^2+3)+3 M^{10}+37 M^8+50 M^6+30 M^4+23 M^2+1\big) \\
    &\qquad -2 M (M^2-1) \dd{m} \big(16 m^4 (M^2+1)^2-8 m^2 (M^2-1)^2 (M^2+1)+(M^4+4 M^2+1)^2\big)\Big), \\
    \bar{\omega}_6 &= Q_A \Big(8 M^2 (M^2-1) \dd{m}-4 M \dd{M} (-4 m^3+3 m M^2+m)\Big), \\
    \bar{\omega}_7 &= Q_A \Big(\dd{m} \big(-16 m^4 (M^4-1)+8 m^2 (M^6-1)-M^8-4 M^6+4 M^2+1\big) \\
    &\qquad +m M \dd{M} \big(16 m^4 (M^2+1)-8 m^2 (M^4+3 M^2+3)+M^6+9 M^4+9 M^2+5\big)\Big), \\
    \bar{\omega}_8 &= Q_A \Big((M^2-1) \dd{m} (-4 m^2+M^2+1)+m M \dd{M} (4 m^2-M^2-3)\Big), \\
    \bar{\omega}_9 &= Q_B \Big(2 M (M^2-1) \dd{m}+m \dd{M} (4 m^2-3 M^2-1)\Big).
  \end{aligned}
\end{equation}
\normalsize